\documentclass[conference]{IEEEtran}
\usepackage{cite}
\usepackage{amsmath,amssymb,amsfonts}
\usepackage{algorithmic}
\usepackage{graphicx}
\usepackage{subfig}
\usepackage{textcomp}
\usepackage[inline]{enumitem}
\usepackage{tabularx}
\usepackage{balance}
\usepackage{color}
\usepackage{placeins}
\usepackage{makecell}
\def\BibTeX{{\rm B\kern-.05em{\sc i\kern-.025em b}\kern-.08em
    T\kern-.1667em\lower.7ex\hbox{E}\kern-.125emX}}

\usepackage[colorlinks=true,linkcolor=black,anchorcolor=black,citecolor=black,filecolor=black,menucolor=black,runcolor=black,urlcolor=black]{hyperref}

\usepackage{xcolor}  % Colored text etc.
\definecolor{OwnAzure}{HTML}{336699}
\definecolor{OwnCerulean}{HTML}{CAE2FE}
\definecolor{OwnOliveGreen}{HTML}{556B2F}
\usepackage[colorinlistoftodos,prependcaption,textsize=tiny]{todonotes}% Add ,disable to the options, to hide the comments
\usepackage{xargs}                      % Use more than one optional parameter in a new commands
\newcommandx{\todoai}[2][1=]{\todo[inline,linecolor=OwnAzure,backgroundcolor=OwnCerulean,bordercolor=OwnAzure,#1]{#2}}
\newcommandx{\addref}[2][1=]
{\todo[inline,linecolor=blue,backgroundcolor=blue!50,bordercolor=blue,#1]{Add reference. #2}}
\newcommandx{\unsure}[2][1=]{\todo[inline, linecolor=red,backgroundcolor=red!25,bordercolor=red,#1]{#2}}
\newcommandx{\change}[2][1=]{\todo[inline, linecolor=blue,backgroundcolor=blue!25,bordercolor=blue,#1]{#2}}
\newcommandx{\info}[2][1=]{\todo[linecolor=OwnOliveGreen,backgroundcolor=OwnOliveGreen!25,bordercolor=OwnOliveGreen,#1]{#2}}
\newcommandx{\improvement}[2][1=]{\todo[linecolor=Plum,backgroundcolor=Plum!25,bordercolor=Plum,#1]{#2}}
\newcommandx{\thiswillnotshow}[2][1=]{\todo[disable,#1]{#2}}

\definecolor{darkred}{rgb}{0.5,0,0}
\definecolor{darkgreen}{rgb}{0,0.5,0}
\definecolor{darkblue}{rgb}{0,0,0.5}

\newcommandx{\wlaskalon}{Askalon}
\newcommandx{\wlchronos}{Chronos}

\usepackage{tabularx}
\newcommand{\multicomment}[1]{}

\usepackage{booktabs} % For formal tables
\newcommand{\ra}[1]{\renewcommand{\arraystretch}{#1}}

\newcommandx{\reviewid}[1]{\todo[linecolor=black,backgroundcolor=orange!25]{#1}}

\begin{document}

\title{A Reference Architecture for Datacenter Scheduling: Design, Validation, and Experiments}

\author{%
[Technical Report on the SC18 homonym article]\\\ \\
 	\IEEEauthorblockN{
         Georgios Andreadis\IEEEauthorrefmark{1}, 
         Laurens Versluis\IEEEauthorrefmark{3},
         Fabian Mastenbroek\IEEEauthorrefmark{1}, and
        Alexandru Iosup\IEEEauthorrefmark{1}\IEEEauthorrefmark{3}}
 	\IEEEauthorblockA{\IEEEauthorrefmark{1}Delft University of Technology, \IEEEauthorrefmark{3}Vrije Universiteit Amsterdam, the Netherlands}
     \{G.Andreadis, L.Versluis, F.Mastenbroek, A.Iosup\}@atlarge-research.com
} %\author

\maketitle

\begin{abstract}
Datacenters act as cloud-infrastructure to stakeholders across industry, government, and academia. 
To meet growing demand yet operate efficiently, datacenter operators employ increasingly more sophisticated scheduling systems, mechanisms, and policies. 
Although many scheduling techniques already exist, relatively little research has gone into the abstraction of the scheduling process itself, hampering design, tuning, and comparison of existing techniques. 
In this work, we propose a reference architecture for datacenter schedulers. 
The architecture follows five design principles: components with clearly distinct responsibilities, grouping of related components where possible, separation of mechanism from policy, scheduling as complex workflow, and hierarchical multi-scheduler structure. 
To demonstrate the validity of the reference architecture, we map to it state-of-the-art datacenter schedulers.
We find scheduler-stages are commonly underspecified in peer-reviewed publications. 
Through trace-based simulation and real-world experiments, we show underspecification of scheduler-stages can lead to significant variations in performance.
\end{abstract}

\vspace*{-0.25cm}

\section{Introduction}
\label{sec:introduction}

Datacenter infrastructure is important for the digital society. Stakeholders across industry, government, and academia employ 
diverse cloud services hosted by datacenter infrastructure, and expect services to be reliable, high speed, and low cost~\cite{tr:idc14cloud,tr:gartner14cloud}. In turn, datacenter operators must maintain efficient operation at unprecedented scale~\cite{book/BeyerJPM16}.
Key to datacenter operation is its {\it scheduler}, which takes on behalf of users and datacenter-engineers decisions about workload and resources~\cite{Ousterhout2013,Verma2015}. 
To keep up with growing demand {and increasing} complexity, 
architects of datacenter schedulers must address
complex challenges in distributed systems~\cite{Tannenbaum2002,Boutin2014,Isard2007a,Verma2015}, software engineering~\cite{DBLP:journals/concurrency/ThainTL05,Hindman2011,DBLP:journals/fgcs/JararwehADBVR16}, and performance engineering~\cite{Ghanbari2012,DBLP:journals/cacm/BurnsGOBW16,DBLP:conf/wosp/IlyushkinAHPGEI17}.
The entry level is high. 
Although {\it reference architectures}~\cite{DBLP:journals/ijhpca/FosterKT01,NISTCloudRA,NISTBigDataRA} help with complexity and entry-level problems,
such {\it conceptual models} currently do not exist for this field.
In this work, we propose a reference architecture for datacenter scheduling, and use it to analyze academia- and industry-designed schedulers.

A conceptual model capturing the entire process of datacenter scheduling could be beneficial to understand how to design, build, and control such complex systems~\cite{DBLP:conf/icdcs/IosupUVAEHTBT18}.
It could facilitate an understanding of how scheduling works in datacenter ecosystems, including the complex and dynamic interplay between the tens of algorithms (policies) currently at work in real deployments. 
The model and associated tools, such as community-curated experimental platforms, could help analyze and compare schedulers in detail, and thus 
counter the problem that many new schedulers and policies (algorithms) appear each year, but few are well-understood or adopted in practice~\cite{DBLP:conf/europar/KlusacekT14}.
By summarizing
patterns and best-practices, the model could
help standardize scheduler construction.

The lack of a reference architecture, or a community-wide underlying conceptual model, can be costly.
Conceptually, this could mean overlooking important components (e.g., replication, see results in Section~\ref{sec:validation:publication}). 
In practice, without a reference architecture, even capable engineering teams could tinker, 
leading to architectural and deployment issues, and 
to challenges in rolling back new projects into the main design.
Condor~\cite{Tannenbaum2002,Thain2005} undertook a successful, but nearly complete and thus time-consuming, redesign of its resource management and scheduling software (around 2006).
Google's Borg had various off-shoot projects, later rolled-back into the main design~\cite{DBLP:journals/cacm/BurnsGOBW16} (e.g., Omega).
As outlined in our vision article~\cite{DBLP:conf/icdcs/IosupUVAEHTBT18}, we aim to understand how to design, build, and control such complex systems.

\begin{figure}[t]
	\centering
    \vspace*{-0.35cm}
    \includegraphics[width=0.8\columnwidth]{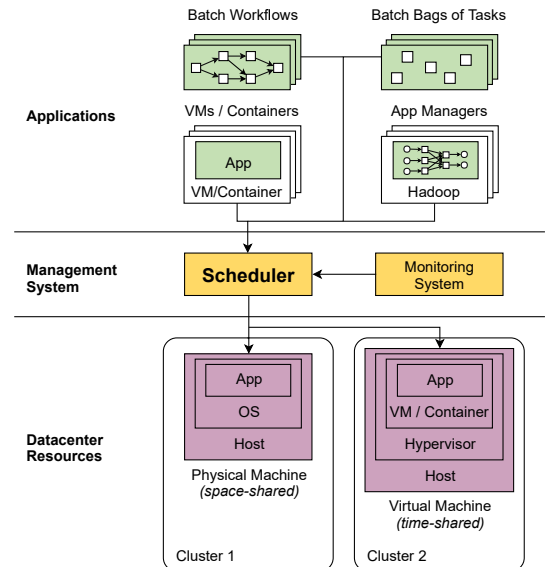}
    \vspace*{-0.15cm}
    \caption{Generic system model for datacenter operation. In this work, we create a reference architecture for the ``Scheduler''.}
    \label{fig:background}
    \vspace*{-0.75cm}
\end{figure}

{\it How to design a reference architecture for datacenter scheduling?} is the fundamental research question addressed in this work.
The answer 
is non-trivial. 
The conceptual model
must exceed the limits of current `black-box' models, because the diversity of actions a scheduler needs to perform is not properly represented by a monolithic and opaque component.
It
must also avoid including too much detail, and thus must trade-off the detail of representation vs. the difficulty of describing the scheduling process; otherwise, the model would become a poor instrument for science, design, and engineering. 
The model must support the main workload, resource, and resource management and scheduling elements
(see Figure~\ref{fig:background}, with terms defined in Section~\ref{sec:model}).

We envision a {\it workflow-based scheduling model}, providing the overall view and a level of detail below it, and the key data and control flows between its workflow-stages.
Inspired by the work of Schopf~\cite{Schopf2004}, who proposed an 11-step abstraction for grid scheduling, we focus on a conceptual model supporting brokered and hierarchical schedulers, but also the diverse set of scheduling operations in datacenters that appears in peer-reviewed research and in practice.
The model supports common datacenter workloads, from  workflows and bags-of-tasks to VMs and container-based hosting of black-box applications, and encompasses common actions in datacenter scheduling, ranging from filtering resources available to the user to task migration. 

To validate the model in terms of validity and usefulness (defined in Section~\ref{sec:ra:req}), we focus on the question of how {\it underspecified} existing schedulers are relative to the conceptual model, that is, how  peer-reviewed specifications about schedulers lack important details for understanding the scheduling-stages. Underspecification is the consequence of a trade-off between the benefits and the costs of documentation. 
Advocates for reproducibility already argue for the importance of specifying designs and experimental setups, both as a principle and in practice~\cite{DBLP:conf/sc/HoeflerB15}. 

The model we envision aligns well with the general benefits of conceptual models, but also enables numerous pragmatic possibilities, two of which we explore in this work.
Through {\it static analysis}, we identify the schedulers with comprehensive detail or underspecification, the scheduling-stages that are least and most underspecified, whether the academia or the industry underspecify the most, whether traditional or recent systems are the most underspecified, etc.
We further use {\it dynamic analysis} to understand the effect of underspecification on the performance and operation of datacenter schedulers at run-time. To this end, we conduct simulation-based and real-world experiments, using as input representative traces collected from real-world environments~\cite{DBLP:journals/fgcs/IosupLJADWE08}.

Overall, the main contribution of this work is three-fold:
\begin{enumerate}
    \item We design a reference architecture for datacenter schedulers (Section \ref{sec:reference-architecture}).
    \item We show how state-of-the-art schedulers can be mapped to the reference architecture, and conduct a static analysis of scheduler specifications (Section \ref{sec:validation}). 
    \item Through dynamic analysis, we assess the impact of underspecification on the performance and operation of datacenter schedulers (Section \ref{sec:experiments}).
\end{enumerate}

\section{Generic Model of Datacenter Operation}
\label{sec:background}\label{sec:model}

We use in this work a generic model of datacenter scheduling that is already widely used in the academia and deployed by the industry~\cite{VanBeek2015,Ousterhout2013,Verma2015}.
Figure~\ref{fig:background} depicts this model.

\subsection{Workload}\label{sec:model:wl}

The workload is comprised of a {\it stream of jobs} of various morphologies, that is, shape, structure, scale, etc. We assume all jobs fit the morphology of {\it workflows}~\cite{Altintas2004,Deelman2016,Kurt2014}: each job consists of a set of one or several \emph{tasks}, with precedence constraints between tasks determining the order of execution. Workflows are common in practice,
from science~\cite[p.137-146]{HeyTT09}\cite{DBLP:journals/ijhpca/DeelmanPACDMPRT18} to engineering~\cite{Wieczorek2005}, from business processes~\cite{DBLP:books/mit/RAH2016} to applications built with serverless cloud services. 

This model,
of {\it workloads of workflows} encompasses:
\begin{enumerate*}
	\item {\it Bag-of-Tasks (BoT)}--jobs comprised of (many) tasks without interdependencies,
    \item {\it hosted jobs}, such as business-critical workloads running inside hosts (e.g., VMs or containers), whose hosts correspond to the long-running tasks of a trivial workflow,
    \item {\it managed jobs}, such as Hadoop- or Spark-based big data applications, or elastic jobs coordinated by an autoscaler~\cite{DBLP:conf/sc/LiuW15}, whose tasks (e.g., Map and Reduce for Hadoop) are coordinated by a long-running master-task in the workflow.
\end{enumerate*}

Workflows can form {\it hierarchical workloads}, that is, tasks in higher-level workflows can themselves take the morphology of a workflow; this process can be recursive.

The user can specify {\it requirements for each job}, which schedulers must fulfill.
To become valid hosts for the job, resources must match {\it resource-requirements}, such as hardware architecture, OS, etc.
Requirements can also include more diverse {\it Service Level Agreements (SLAs)}~\cite{DBLP:conf/icws/ComuzziKSY09}, such as the cost of leasing the resource~\cite{DBLP:conf/sc/DeelmanSLBG08,DBLP:journals/sp/JacksonMRRT11}, elasticity~\cite{DBLP:conf/icac/HerbstKR13}, licenses and other codified legal aspects, etc.

\subsection{Datacenter Resources}\label{sec:model:dc}

In our model, datacenters are comprised of physical and virtual resources.
We consider a generic resource model: physical resources, such as nodes in a datacenter rack, and virtual resources, such as VMs or containers, are typically grouped into physical or virtual \emph{clusters}~\cite{Verma2015}.

We model in this work resources only through abstract operation, a model commonly employed in analytical and simulation-based studies~\cite{DBLP:journals/jpdc/CasanovaGLQS14,DBLP:journals/spe/CalheirosRBRB11}: we model (1) {\it processing resources} with a generic model of production, e.g., FLOPS or MIPS, (2) {\it memory and storage resources} with a generic model of size and bandwidth but no latency, etc.

\subsection{Resource Management and Scheduling}\label{sec:model:scheduling}

The {\it user} submits a workload to the datacenter, through a {\it central submission site}. 
Serving the central submission site is the \emph{scheduler}, which follows a common but complex operational model~\cite{Ousterhout2013,Verma2015}: the scheduler matches streaming workloads to existing resources, selects and allocates resources for jobs, conducts setup actions for jobs and tasks, executes the workload, and also manages the workload through its life-cycle~\cite{Verma2015} (e.g., from submission and setup to completion and cleanup).
If an application requires more resources during its execution, or can release some of the resources it has been allocated, the scheduler can use an autoscaler to conduct these dynamic, elasticity-related {\it provisioning operations}. Autoscalers can be workflow-agnostic~\cite{Ghanbari2012} or workflow-specific~\cite{DBLP:conf/wosp/IlyushkinAHPGEI17}.

While executing the workload, the scheduler may preempt, recover, or migrate tasks, based on various events. Events may include performance variability~\cite{DBLP:journals/simpra/MathaRP17}, (correlated) resource failure~\cite{DBLP:conf/sc/HeienKGLKC11}, evolving leasing costs~\cite{DBLP:journals/fgcs/JavadiTB13}, etc.
Events are provided primarily by the ``System monitor'' component (Figure~\ref{fig:background}), which continuously gathers resource-state information.

Although schedulers can operate as central monoliths, they can also be divided into smaller, cooperating instances.
This can be the case in {\it hierarchical scheduler} models, where the scheduler operates as a hierarchy that typically matches the datacenter topology~\cite{Iosup2011}. For example, local schedulers (\emph{Local Resource Manager (LRM)}) could manage each a single cluster, with a datacenter-level scheduler distributing workload between LRMs.
Custom setups that deviate from the standard scheduling flow can be facilitated by a \emph{broker}~\cite{Pawluk2012}, which is a component that can negotiate resources for workloads directly, potentially across different providers (schedulers).

The scheduler takes decisions guided by {\it policies}, separated by design from the mechanism that executes the decisions. Many policies already exist for scheduling in datacenters and clouds~\cite{Singh2016,Rodriguez2017}.
For example, 
to rank the tasks eligible for execution, 
schedulers could use Shortest-Remaining-Time-First ({\tt SRTF}), where tasks with shorter (estimated) execution time are given a higher priority over tasks with longer times.
As another example, to take decisions related to resource allocation, schedulers could use the {\it First-Fit} policy, which selects the first resource with enough capacity to fit the task.

We explicitly do not consider here the scheduling processes occurring at user- and/or framework-level, that is, dynamically managing the computation and communication processes occurring inside the application. For example, the general model supports the workflows that users would build in Spark, but not the fine-grained, framework-level operations, such as RDD checkpointing, migration, and broadcasting.

\begin{figure}
	\includegraphics[width=\linewidth]{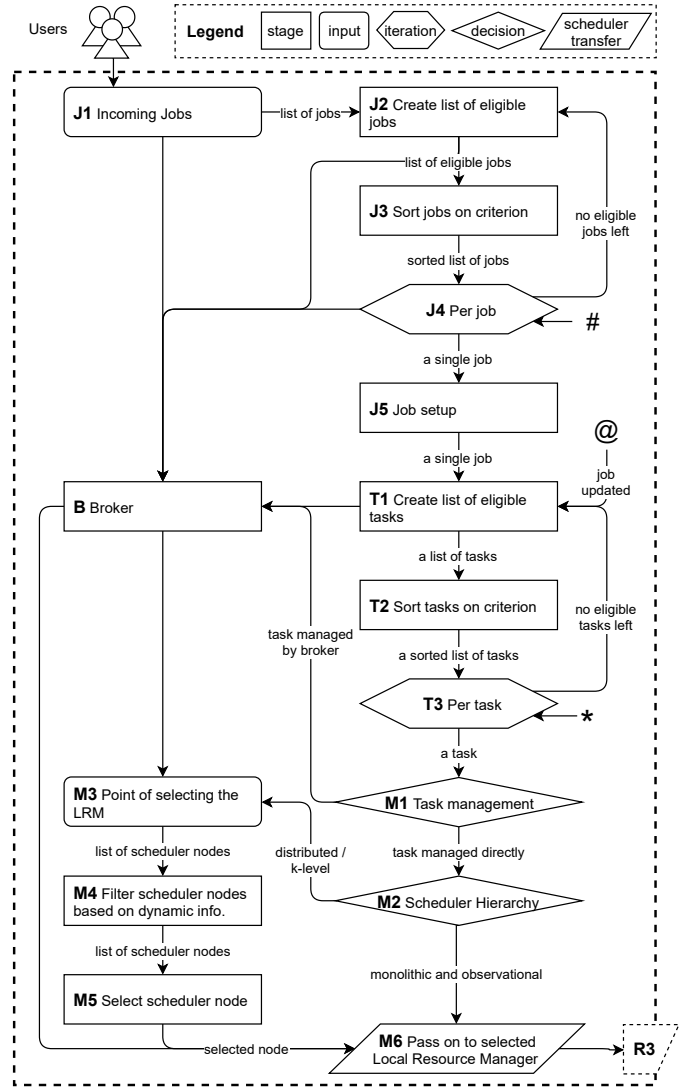}
    \caption{Reference architecture for datacenter scheduling. Focus on the {\it global} scheduler. (Continued in Figure~\ref{fig:scheduling-flow-2}: the *, @, and \# symbols are jump labels to Figure~\ref{fig:scheduling-flow-2}.)}
    \label{fig:scheduling-flow-1}
    \vspace*{-0.5cm}
\end{figure}

\section{Design of a Reference Architecture for Datacenter Scheduling}
\label{sec:reference-architecture}
\label{sec:ra}

\begin{figure*}
   \centering
	\includegraphics[width=0.85\linewidth]{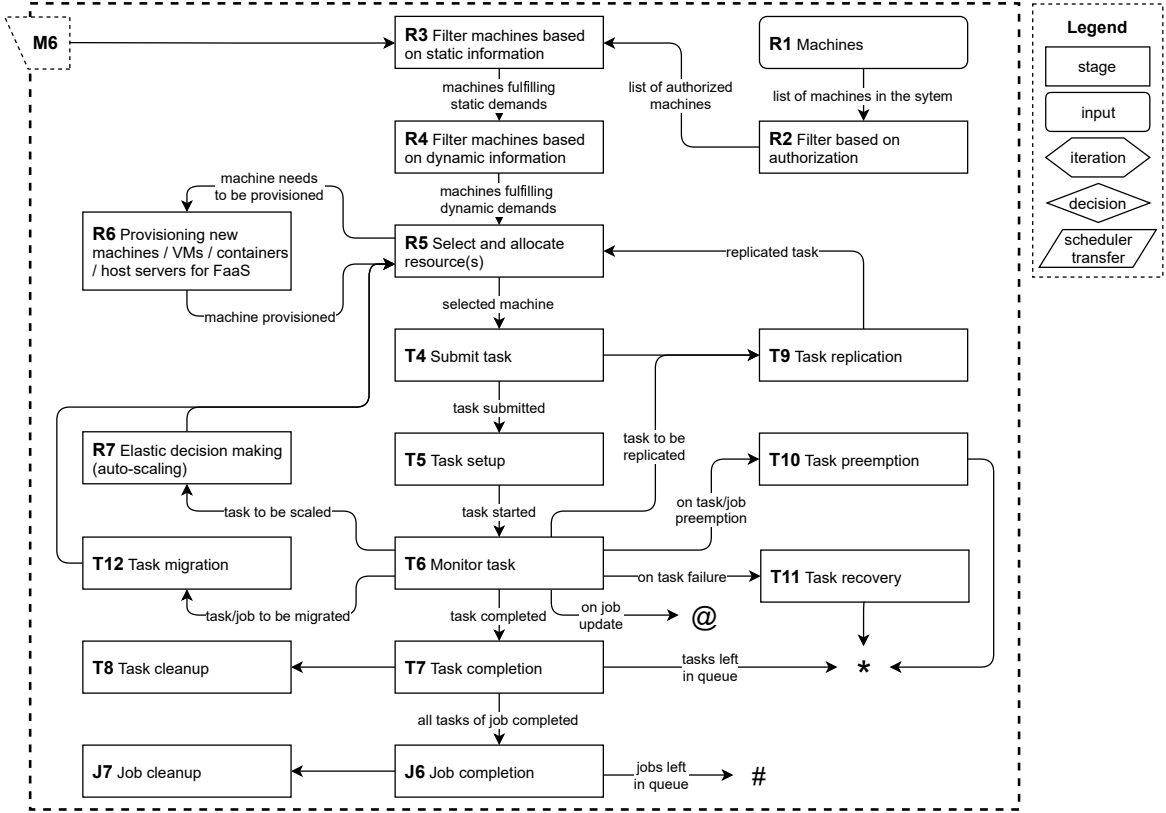}
    \caption{Reference architecture for datacenter scheduling. Focus on the {\it local} scheduler. (Continued from Figure~\ref{fig:scheduling-flow-1}.)} %: the *, @, and \# symbols are jump labels to Figure~\ref{fig:scheduling-flow-1}.)}
    \label{fig:scheduling-flow-2}
    \vspace*{-0.5cm}
\end{figure*}

In this section we design the reference architecture. 
We discuss the main requirements and the design principles that influenced our design process. 
We then propose a structured, workflow-based approach to scheduling, whose core data and control flows we depict in Figures~\ref{fig:scheduling-flow-1} and~\ref{fig:scheduling-flow-2}.

\subsection{Main Requirements} \label{sec:ra:req}

Every design is created to fulfill a set of requirements, corresponding to the key problems identified by the design stakeholders. Key stakeholders to this design are researchers in the field of resource management and scheduling, datacenter operators, and students interested in the field. In this work, we consider two main requirements, validity and usefulness, which correspond to the complete set of stakeholders.

{\it R1. Validity} is the property of the proposed model to accurately represent the field of datacenter scheduling.
For this, the model needs to cover the state-of-the-art, and in particular instances of datacenter scheduling from both real-world practice and emerging concepts from the academia.
Although verifying this requirement is fundamentally a subjective task, mapping existing schedulers to the model embodied by the reference architecture (discussed in Section~\ref{sec:validation}) {and peer-review} both give evidence the model is relevant.

{\it R2. Usefulness} gives the reference architecture a real-world purpose, which motivates its creation.
As with validity, this property cannot be measured exactly, but the empirical results presented in Sections~\ref{sec:validation} and~\ref{sec:experiments}, combined with the future uses of this architecture we envision in Section~\ref{sec:conclusion}, give evidence of usefulness for the reference architecture.

\subsection{Leading Design Principles} \label{sec:ra:principles}

We have derived five leading principles for the design of the reference architecture for datacenter scheduling, which we use to design the reference architecture in Sections~\ref{sec:ra:overview} and~\ref{sec:ra:details}:

\emph{P1. Components with Clearly Distinct Responsibilities:}
Essential to our design is the decomposition of the datacenter-scheduling process into individual components.
Derived from the common practice of designing software architectures~\cite{Rozanski2005}, each system component has a set of responsibilities, defined boundaries of those responsibilities, and a set of interfaces which define its services to other components.
The set of components identified in the design-process gives choices when designing specific schedulers---not all components are used by every datacenter-scheduler deployed in practice.

\emph{P2. Grouping of Related Components:}
Corresponding to best-practices for packaging components~\cite{Rozanski2005}, related components can be grouped according to their responsibility. 
Depending on stakeholder and/or viewpoint, the components of the same architecture can lead to different grouping schemes.

\emph{P3. Separation of Mechanism from Policy:}
The definition of abstract components separates mechanism from policy. 
To provide the necessary level of abstraction, the reference architecture should not mandate specific policies---this model is \emph{qualitative}, as defined in~\cite{Rozanski2005}.

\emph{P4. Scheduling as Complex Workflow, Matching the System Model} (see model defined in Section~\ref{sec:model:scheduling}):
Schedulers act as workflows, with well-defined data and control flows between the workflow components. Although we consider in this work reference data- and control-flows, we can expect that in practice different flows will emerge, related to the physical topology and the operational goals of each datacenter.

\emph{P5. Hierarchical Scheduler with Shared Control:}
Derived from the system model, the scheduler is structured as a hierarchy, subject to the central control of the datacenter operator, and to partial control by or on behalf of individual datacenter-users. Both the hierarchy and the model for shared control are expressed through policies, but the latter policies may be changed more frequently in practice.

We have been able so far to map many complex schedulers to the reference architecture built with these principles (next section).
Through induction, we argue the principles and practice of this process can continue to succeed, subject to the slow-paced evolution of the reference architecture.

\subsection{Overview of the Reference Architecture} \label{sec:ra:overview}

We present in this section the {\it overall design} and the {\it grouping} of components addressing the major responsibilities of datacenter scheduling.

Following principle {\it P1}, the reference architecture for datacenter scheduling depicted in Figures~\ref{fig:scheduling-flow-1} and~\ref{fig:scheduling-flow-2} models the scheduler as a set of components (term used interchangeably in this work with {\it stages}), each with specified inputs, outputs, and side-effects, the combination of which define its function. Each stage can be equipped with its own policy ({\it P3}), which is often a simple, heuristic algorithm. %We detail this part in 
Section~\ref{sec:ra:details} details this.

Following {\it P4}, the scheduler forms, through control and data flows, a workflow that matches the main functions defined in the system model (Section~\ref{sec:model:scheduling}):  execution starts at the central submission site, and progresses until job and task completion and cleanup stages. The execution flow of the scheduler (one {\it scheduler iteration}) may be started periodically, on events such as job arrival or completion, or even manually. In the scheduling workflow, control is transferred between stages through outputs, which are depicted as text on the interconnections between components in Figures~\ref{fig:scheduling-flow-1} and~\ref{fig:scheduling-flow-2} (e.g., in Figure~\ref{fig:scheduling-flow-1}, stage {\tt J1} transfers data, a ``list of jobs'', to stage {\tt J2}), or through events such as the completion of a stage, which are indicated as interconnections without description in the same figures (e.g., in Figure~\ref{fig:scheduling-flow-2}, the completion of stage {\tt T7} triggers the start of stage {\tt T8}). 

Following {\it P2}, we identify four major responsibilities: %we group the components into four major responsibilities: 
\begin{enumerate}
\item {\it Job processing} (\texttt{J}): activities concerning the selection of jobs and the job life-cycle, such as job setup and cleanup,

\item {\it Task processing} (\texttt{T}): stages of the task life-cycle, including more sophisticated stages for task migration, preemption, and replication,

\item {\it Scheduler management} (\texttt{M}): stages facilitating scheduling hierarchies and (cloud) brokers, and

\item {\it Resource management} (\texttt{R}): stages related to provisioning and allocating resources.
\end{enumerate}

Following {\it P2} and {\it P5}, we define another grouping of stages to emphasize the scheduler hierarchy: Figure~\ref{fig:scheduling-flow-1} depicts the stages executed by a {\it global} scheduler (datacenter-level), whereas Figure~\ref{fig:scheduling-flow-2} visualizes stages predominantly executed at a subordinate, {\it local} scheduler (cluster-level). 
As for control and data flow, this categorization only serves as a suggestion and may be deviated from for different schedulers (e.g., a monolith scheduler may execute all stages on the same machine).
The {\it Broker} component is represented as a stub; the broker itself can employ more complex sub-systems, in a design that remains external to this architecture.

\subsection{Details of the Reference Architecture} \label{sec:ra:details}

In this section, we present the design of the individual stages addressing the major responsibilities of datacenter scheduling.

To design the reference architecture at this level of detail,
we focus each stage on a specific responsibility, and select for it one of the five {\it operational modes}: processing, input, iteration, decision, and transfer.
For example, {\it iterative stages}, such as \texttt{J4} and \texttt{T3} in Figure~\ref{fig:scheduling-flow-1}, repeat the remainder of the flow they are connected to, for each element of input that they receive.
However, the execution of these iterations is not restricted to sequential execution.
Iterative stages may launch multiple, parallel threads of execution, that is, \texttt{J4} may trigger multiple jobs to be processed in parallel, each starting with \texttt{J5} and going through the rest of the process, independently.
(This requires parallel-code constructs and consistency protocols, including {\it Paxos}~\cite{DBLP:journals/cacm/BurnsGOBW16}.)

This reference architecture also permits stages to be either {\it stateless} or {\it stateful}. Stateful stages persist state across invocations and take historical information into account.
We further differentiate between two types of {\it expected outputs} that these stages can produce: 
{\it outputs as a function of input} (including historical data for stateful stages), such as the set outputs of task filtering sorting stages, and {\it side-effects}, such as the actions of job setup and cleanup stages. 

Below follows a full description of all stages:

\texttt{J1} -- \emph{Incoming jobs}:
this input stage provides the list of jobs that users have submitted at the central submission site of the scheduling system.

\texttt{J2} -- \emph{Create list of eligible jobs}:
this stage makes a selection from the input list of jobs that only includes jobs that are eligible to be scheduled. 
A policy chosen for this stage may dictate that all jobs may be passed through, or may implement a restriction on eligibility, e.g. due to certain user restrictions or flow control measures. 
Any jobs rejected at this stage return to the list  of incoming jobs and are reconsidered at the next scheduling iteration.

\texttt{J3} -- \emph{Sort jobs on criterion}:
this stage sorts a list of jobs based on a certain priority criterion and outputs the sorted list. 
Policies that determine this priority can take a variety of meta data into account, such as: the submitting user, a metric such as estimated time of completion, or even a composite score of different aspects.

\texttt{J4} -- \emph{Per job}:
this stage passes each job to the rest of the pipeline and remains in control until all jobs have been passed on. 
It therefore acts as a branching point for the flow of execution.
In practice, this component can also trigger multiple parallel runs of the pipeline.

\texttt{J5} -- \emph{Job setup}:
this stage takes a job and performs any setup actions the job needs to perform before its tasks commence. 
An example of such an action is a data transfer of required files to the resource site.

\texttt{J6} -- \emph{Job completion}:
this stage is entered on completion of a job. 
If needed, this stage can notify the submitting user of completion and execute job-level data transfers back to the submitting user.

\texttt{J7} -- \emph{Job cleanup}:
this stage is entered after completion of a job and is responsible for job-level cleanup actions necessary after the execution of a job (e.g. deletion of data files). 

\texttt{T1} -- \emph{Create list of eligible tasks}:
this stage filters the list of tasks provided as input, based on a filter-policy, e.g. a policy that allows tasks to pass through if and only if their dependencies have already finished.

\texttt{T2} -- \emph{Sort tasks on criterion}:
this stage takes a list of tasks and sorts it on a given criterion.
This can be done to improve the throughput and latency of tasks.
Policies here include {\tt SRTF}, {\tt FIFO}, and even uniformly random ({\tt RANDOM}).

\texttt{T3} -- \emph{Per task}:
this stage is the task-level equivalent to \texttt{J4}.
It operates in the same fashion, but handles tasks instead of jobs.

\texttt{T4} -- \emph{Submit task}:
this stage submits the given task to the resource that is provided as input.
This is a final reservation of the portion of the resource that the task will occupy.

\texttt{T5} -- \emph{Task setup}:
this stage transfers any executable or data files required by the task to its machine.
Once this transfer has been completed, the task's main executable is launched.

\texttt{T6} -- \emph{Monitor task}:
this stage periodically receives monitoring information from the monitoring system on the progress of the task towards completion and resource usage. 
Policies implemented here could decide to preempt a task or to start recovery actions upon failure, based on information provided by the monitor.

\texttt{T7} -- \emph{Task completion}:
this stage is triggered when the task is identified as completed and marks it as such in its internal records.

\texttt{T8} -- \emph{Task cleanup}:
this stage is entered after the completion of a task and can be used for task-level cleanup actions, such as the deletion of data files and executables.

\texttt{T9} -- \emph{Task replication}:
this stage replicates the task currently in the pipeline and passes the replicated copy (or copies) back to the resource selection stage (\texttt{R5}).
This can enhance the persistence of the system against failures.

\texttt{T10} -- \emph{Task preemption}:
this stage is entered when a task or job is ordered to be preempted.
This can be initiated by a manual user or operator event, or an automated scheduler decision.
On preemption, task execution is aborted and the task is sent back into the task queue.

\texttt{T11} -- \emph{Task recovery}:
this stage can be triggered by task failure and takes appropriate recovery measures to address this failure.
After any recovery tasks (e.g. sending back partial results) have been performed, the task is placed in the task queue again, where the task can later be rescheduled.

\texttt{T12} -- \emph{Task migration}:
this stage can trigger the migration of a task to a different resource.
Such a migration can be initiated manually or by the scheduling system, depending on the chosen policy.

\texttt{M1} -- \emph{Task management}:
this stage passes control to the broker, if and only if such a broker is present and the stage deems this task to be under control of the broker.

\texttt{M2} -- \emph{Scheduler hierarchy}:
depending on whether the scheduler is operating in a hierarchy, this stage passes control to a next stage.
The proposed model has two possible modes of operation: distributed and monolithic.

\texttt{M3} -- \emph{Point of selecting the LRM}:
this stage provides a list of scheduler nodes connected to the global scheduler.
Note that this list includes both available and unavailable nodes -- the latter are filtered out in stage \texttt{M4}.

\texttt{M4} -- \emph{Filter scheduler nodes based on dynamic information}:
this stage filters the scheduler nodes based on their availability.
Policies for this may include the number of tasks currently managed by that node or the fraction of computing cores still available in the resources managed by that node.

\texttt{M5} -- \emph{Select scheduler node}:
this stage makes the decision of which scheduler node to allocate the given task to.
The operation of this stage closely resembles the \texttt{R5} stage, with the difference that this stage operates on scheduler nodes instead of machines.

\texttt{M6} -- \emph{Pass on to selected LRM}:
in this stage, control is transferred to the selected scheduler node.
If the system has a monolithic scheduling structure, this stage has no effect on the task.
If the system has a distributed scheduling hierarchy, the task is passed on to the selected scheduler node and that node resumes control from this point on.
Stages connected by outgoing links to this stage are therefore executed on that scheduler node.

\texttt{B} -- \emph{Broker}:
the cloud broker acts as an alternative resource management entity, facilitating custom control over the workload and resources.
Implementations for this component can vary significantly in scope and complexity.

\texttt{R1} -- \emph{Machines}:
this input stage provides a list of resources to the system.
This information is typically based on static architectural knowledge of the datacenter.
This stage returns both available and offline machines, the latter of which can be filtered out in later stages.

\texttt{R2} -- \emph{Filter based on authorization}:
this stage filters the list of machines provided to it on authorization constraints.
Such constraints may dictate that certain customers do not have access to a given subset of machines.

\texttt{R3} -- \emph{Filter machines based on static information}:
this stage examines the statically verifiable resource requirements of the workload (e.g. the host OS, hardware architecture, etc.) and only passes machines through that conform to these requirements.

\texttt{R4} -- \emph{Filter machines based on dynamic information}:
this stage acts as a filter yielding a list of resources with sufficient resource-capacities, based on fixed or dynamic requirements, and on predicted or monitored  information about processing unit availability, memory occupancy, etc.

\texttt{R5} -- \emph{Select and allocate resource(s)}:
in this stage, the selected task is matched with a (set of) resource(s), using policies such as First-Fit, Worst-Fit, and Best-Fit.
The match is passed on to the task submission stage (\texttt{T4}).

\texttt{R6} -- \emph{Provisioning new machines / VMs / containers / host servers for FaaS}:
this stage allows a managing entity to provision resources.
This may be necessary if the user does not have any machines left at his/her disposal, at which point new resources will need to be reserved to accommodate for the incoming task(s).

\texttt{R7} -- \emph{Elastic decision making (auto-scaling)}: 
this stage provides room for dynamic resource management decisions, scaling the pool of resources available to the user, job, or task.

\section{Validation through Mapping of Schedulers}
\label{sec:validation}

\begin{table}[!t]
	\centering
	\caption{Overview of the mapping-based validation.}
	\label{tab:mapping-overview}
    \vspace*{-0.25cm}
	\begin{tabularx}{\columnwidth}{lXcc}
		\toprule
		Section(s) & Analysis & Scheduler(s) & Aggregation   \\ \midrule
		\ref{sec:validation:borg} & Detailed mapping & Borg & None (all stages) \\
		\ref{sec:validation:overview}  & Mapping overview &  All & By stage-group  \\ 
		\ref{sec:validation:industry-academia}, \ref{sec:validation:publication}  & Head-to-head comp. &  All &  By feature \\ 
        \bottomrule
	\end{tabularx}
    \vspace*{-0.5cm}
\end{table}

Reference architectures are most useful when they accurately depict real-world instances. 
In this section, we validate the reference architecture by mapping to it fourteen diverse real-world, well-known, and state-of-the-art schedulers, in the three experiments using static analysis summarized in Table~\ref{tab:mapping-overview}.

The schedulers are selected based on two criteria: (1) the scheduling core (e.g., policies) have been analyzed and presented for the systems community, and (2) the peer-reviewed material is highly cited or used in a company running large datacenters.
The first, detailed mapping of one scheduler to the reference architecture, is intended to exemplify the mapping process (see Section~\ref{sec:validation:process}).

The second mapping, in Section~\ref{sec:validation:overview}, tests the validity of the reference architecture (requirement {\it R1}), by conducting the detailed mapping of each scheduler considered in this work, then aggregating the results into a tabular overview.
We define {\it underspecification} for a scheduler as the degree of specification, as presented in peer-reviewed publications, of
the scheduling stages included in the reference architecture, and of the policies used by the scheduler.
{\it Which (groups of) stages are the most, or the least, underspecified? Which schedulers do the most, or the least, in specifying their stages?}

Fulfilling requirement {\it R2}, the mappings in Sections~\ref{sec:validation:industry-academia} and~\ref{sec:validation:publication}
help understand how schedulers map to the reference architecture, by asking relevant questions that divide schedulers into feature-based classes: {\it Are schedulers designed in the academia better specified than industry-designed schedulers? Are newer designs better specified than older designs?}

\subsection{The Mapping Process} \label{sec:validation:process}
The mapping process proceeds as follows. 
Two reviewers have split the mapping work, conducting parallel mapping processes except for a small overlapping set of schedulers, including Borg~\cite{Verma2015}, used for calibration; whenever the results for the overlapping set were found to be different across reviewers, they have discussed the reasons, reconciled their views, and adjusted their mapping process accordingly.
Both experts and non-experts can apply this process (see Section~\ref{sec:mapping:repro}).

For each scheduler, we start the mapping process from the relevant peer-reviewed publications summarized in Table~\ref{tab:mapping-table}. We further consider other publications, and \textit{source-code inspection} when implementation details of the schedulers are accessible and well documented, for example through GitHub. 

Stages for which the reviewers can find a credible description of both what the stage does, and how (both mechanism and policy), are labeled as {\it ``full match''}.
Stages for which the reviewers can find only one of the what and how, or for which the description of the how does not credibly match the what, are labeled as {\it ``partial match''}. The remaining stages are labeled {\it ``no match''}. A special case occurs when the scheduler has both a peer-reviewed publication and publicly accessing source-code: we do not make a distinction between stages that are not mentioned in the publication and stages that are not present in the actual scheduler implementation; in particular, the former results in a ``no match'' label for that stage. 

\subsection{Exemplary Mapping of Borg (Google)} \label{sec:validation:borg}

\begin{figure}
\centering
\includegraphics[width = 5cm]{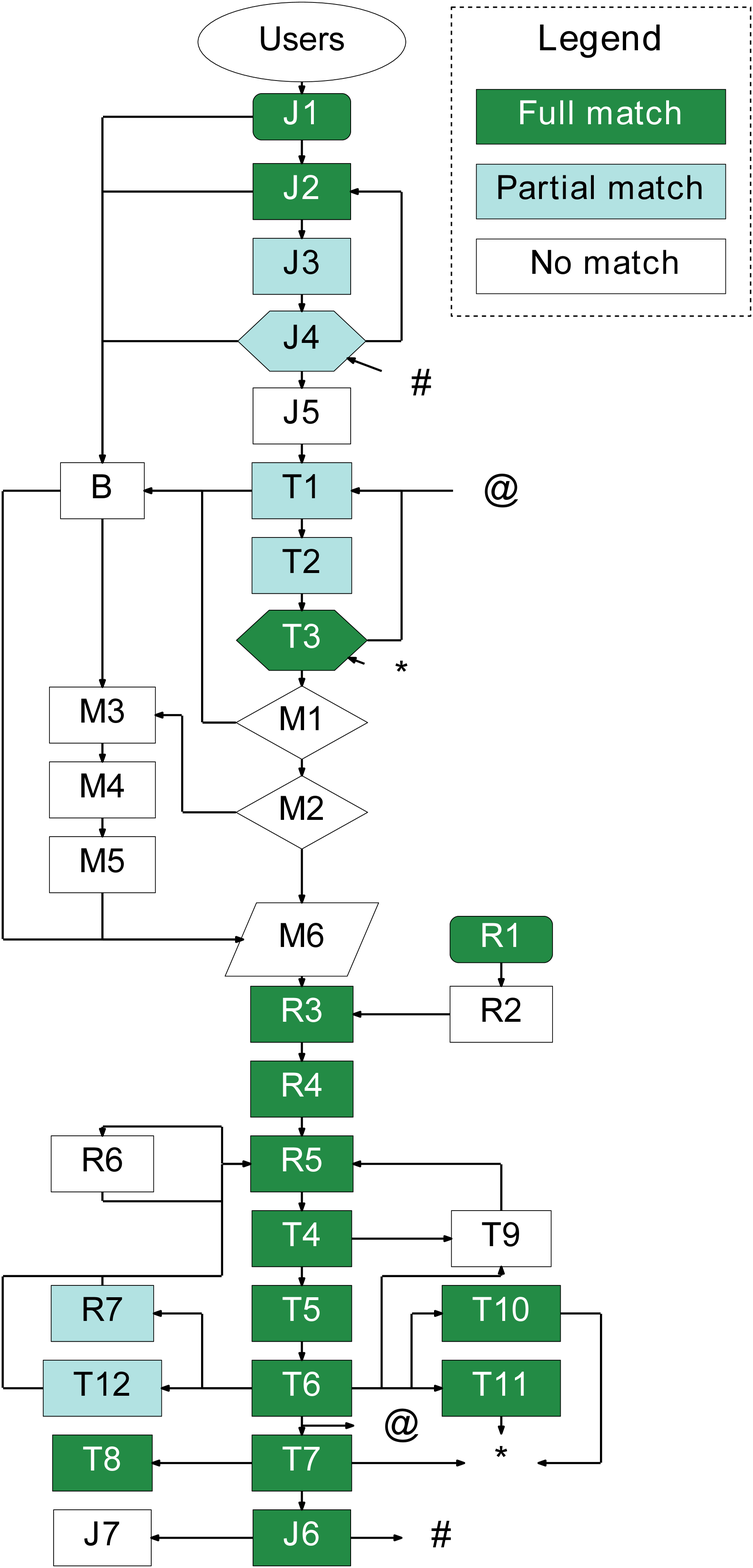}
\vspace*{-0.25cm}
\caption{Borg (Google) mapped to the reference architecture.}
\label{fig:mapping-flow-borg}
\vspace*{-0.5cm}
\end{figure}

To illustrate how the reference architecture allows studying state-of-the-art systems, in this section we map to the reference architecture Google's Borg scheduler~\cite{Verma2015}, and depict the result in Figure~\ref{fig:mapping-flow-borg}. Although scheduling at Google has evolved, Borg remains the seed of the concepts seen later in Omega and in Kubernetes~\cite{Verma2015}, and the learnings of the Omega project have been reabsorbed into the production-scheduler Borg~\cite{DBLP:journals/cacm/BurnsGOBW16}.

\subsubsection{Workload}
Similarly to the proposed reference architecture, users provide the Borg system with a workload in the form of jobs, structured as bags of tasks. 
The Borg system also allows jobs to be prioritized, both coarse-grained (i.e., production or non-production) and more fine-grained. 
This matches the `Sort jobs on criterion' stage (\texttt{J3}), but we label this stage ``partial match'' because, see~\cite[\S2.5]{Verma2015}, (i) the production tasks are allocated a majority of the resources and cannot preempt each other, (ii) the details for fine-grained priority indicate the need for careful manual tuning, and (iii) no details are provided on whether the tasks are actually sorted next to being ranked.
Jobs in Borg can also have constraints on, e.g., processors or OSs they can run on. This matches the `Filter machines based on static information' stage (\texttt{R3}); here, enough information exists to label this stage a ``full match''.

\subsubsection{Resource Organization}
Jobs generally run in one ``cell'', that is, a set of machines that is considered one unit. One or more cells run in a physical cluster, and one or more clusters form a datacenter. Cells are managed by a central scheduling entity (``Borgmaster'', global) and machines by a local agent (``Borglet'', local). 
In this hierarchical model, the global scheduler selects, at the `Scheduler Hierarchy' decision point of the architecture (\texttt{M2}), the lower-level (local) scheduler nodes to take over each job. (For users having access to multiple Borgmasters, we can envision another hierarchical level, whose use is possibly facilitated by the broker (\texttt{B}), but this does not appear in the Borg description~\cite{Verma2015}.)

Corresponding to jobs based on VMs or containers in our model, resources can be reserved through ``allocs'' (single machine) and ``alloc sets'' (multiple machines).  
This corresponds to the `Select and allocate resource(s)' stage (\texttt{R5}).

\subsubsection{Task Life-Cycle and Operations}
All of the {\tt T}-stages apply to the Borg system. For \texttt{T6}, ``Borglets'' assume the role of monitoring tasks, providing information about their machines to higher-level schedulers (and to human engineers). 

The Borg system supports policy-based task replication (\texttt{T9}), preemption (\texttt{T10}), recovery (\texttt{T11}), and for some jobs also migration (\texttt{T12}).
For example, Borg uses for resource reclamation a combination of `Task monitoring' (\texttt{T6}), `Task preemption' (\texttt{T10}), and `Elastic decision making' (\texttt{R7}, but the autoscaling mechanism lacks detail in~\cite[\S5.5]{Verma2015}): to ensure efficient use of all resources, the Borgmaster monitors per-task usage, real-time, and marks unused resources as free for lower-priority jobs (such as batch jobs). 
When a task needs again more resources, lower-priority jobs are preempted.

\subsubsection{Scheduling Algorithm}
At Borgmaster-level, the scheduling process consists of two main steps.
First, during feasibility checking, the process selects machines fitting the task, by combining static and dynamic information, and thus by using both machine-filtering stages (\texttt{R3} and \texttt{R4}). 

Second, corresponding to the `Select and allocation resource(s)' stage (\texttt{R5}), the process gives each machine a score based on a combination of metrics, and then uses a policy to select a machine (e.g., First- and Best-Fit).

\begin{table}[!t]
\centering
\caption{Summary of schedulers mapped to the reference architecture. Numbers express matching per group, as mean percentage over the scheduling tasks of the group, with 100\%/50\%/0\% for a full/partial/no match per task (see text).}
\label{tab:mapping-table}
\vspace*{-0.25cm}
\ra{1.2}
\begin{tabularx}{\columnwidth}{Xcrrrrrrrr} \toprule
& & \multicolumn{4}{c}{Group of Components} & \multicolumn{4}{c}{Selected Components\textsuperscript{$\star$}} \\ 
\cmidrule(r){3-10} Scheduler & Feat.\textsuperscript{$\star$} & \multicolumn{1}{c}{\texttt{J}} & \texttt{T} & \texttt{M} & \texttt{R} & \multicolumn{1}{c}{\texttt{T1}} & \texttt{T2} & \texttt{R4} & \texttt{R5}\\ 
\midrule
Condor~\cite{Tannenbaum2002} & A,O,M & 64 & 88 & 28 & 71 & $\checkmark$ & $\checkmark$ & $\checkmark$ & $\checkmark$\\
Mesos~\cite{Hindman2011} & A,N,M & 14 & 46 & 0 & 71 &  &  & $\checkmark$ & $\checkmark$\\
Borg~\cite{Verma2015} & I,N,M & 57 & 79 & 0 & 64 & $\sim$ & $\sim$ & $\checkmark$ & $\checkmark$\\
Fuxi~\cite{Zhang2014} & I,N,M & 57 & 54 & 0 & 50 & $\sim$ &  & $\checkmark$ & $\sim$\\
Autopilot~\cite{Isard2007} & I,O,S & 36 & 46 & 0 & 29 &  &  &  & $\sim$\\
\midrule
Sparrow~\cite{Ousterhout2013} & A,N,S & 21 & 33 & 57 & 0 &  & $\sim$ &  & \\
Pegasus~\cite{Deelman2016} & A,N,M & 71 & 63 & 0 & 36 & $\checkmark$ &  &  & $\sim$\\
Quincy~\cite{Isard2009} & I,O,S & 43 & 50 & 0 & 28 &  &  &  & $\checkmark$\\
ICENI~\cite{McGough2004} & A,O,S & 50 & 33 & 0 & 57 & $\sim$ &  &  & $\checkmark$\\
Firmament~\cite{Gog2016} & A,N,S & 14 & 29 & 28 & 57 & $\sim$ & $\checkmark$ & $\checkmark$ & $\checkmark$\\
\midrule
Apollo~\cite{Boutin2014} & I,N,S & 85 & 75 & 100 & 57 & $\checkmark$ & $\checkmark$ & $\checkmark$ & $\checkmark$\\
Askalon~\cite{Wieczorek2005} & A,O,M & 71 & 8 & 14 & 14 & $\sim$ &  &  & \\
Triana~\cite{Majithia2004} & A,O,S & 43 & 42 & 7 & 14 & $\checkmark$ & $\checkmark$ &  & $\sim$\\
Dryad~\cite{Isard2007a} & I,O,M & 64 & 71 & 71 & 43 & $\checkmark$ &  & $\sim$ & $\sim$\\
\bottomrule
\multicolumn{10}{p{8cm}}{
($\star$) Features: A/I = Academia/Industry, 
O/N = before/in-or-after 2010, 
S/M = Single-/Multi-Cluster.\newline
Stages: $\checkmark$/$\sim$/space (\textvisiblespace{}) = full/partial/no match. }\\
\end{tabularx}
\vspace*{-0.25cm}
\end{table}

\begin{table}[!t]
	\centering
	\caption{The top-10 stage with the largest difference in match (\%), for systems from academia (A) vs. industry (I).}
	\label{tab:component-differences}
    \vspace*{-0.25cm}
    \ra{0.9}
    \newcolumntype{C}{>{\centering\let\\\arraybackslash\hspace{0pt}}p}
	\begin{tabularx}{\columnwidth}{Xrrrrrrrrrr}
		\toprule
		Field   & \texttt{T11}  & \texttt{T10}  & \texttt{J2}  & \texttt{R2}    & \texttt{J7}    & \texttt{J3}   & \texttt{J6}   & \texttt{T9} & \texttt{R3} & \texttt{T8}	\\ \midrule
		A & 13 & 31 & 6 & 38  & 38  & 6  & 56 & 13 & 38 & 44 \\
		I & 83 & 92 & 50  & 0     & 0     & 42 & 92 & 42 & 67 & 17 \\ \bottomrule
	\end{tabularx}
    \vspace*{-0.5cm}
\end{table}

\subsection{Mapping Overview}
\label{sec:validation:overview}

We have repeated for all schedulers considered in this work the process described in Section~\ref{sec:validation:process} and outlined for Borg in Section~\ref{sec:validation:borg}. 
Table~\ref{tab:mapping-table} summarizes the results of the mapping, including statistics for the {\it groups} of stages for each of the four major responsibilities (see Section~\ref{sec:ra:overview}), and detailed information for four selected stages (the focus of experiments in Section~\ref{sec:experiments}).
For each group, we first rate the stages in the group (``full'', ``partial'', and ``no'' matches receive a rating of 100\%, 50\%, and 0\%, respectively), then compute an aggregate-percentage as the arithmetic mean of the ratings of each stage in the group.

The results indicate the presence of a wide difference, across schedulers, in the level of specification of both grouped and individual stages. The full specification of stages is rare, and only Apollo, Condor, Dryad, and Borg specify over 50\% of their components. Conversely, the most underspecified group is ``scheduler management'' (less than half of schedulers specify anything about {\tt M}-components) and ``resource management'' (no scheduler defines everything, or even three-quarters of the {\tt R}-components). The most underspecified schedulers include Triana, Sparrow, and Autopilot. Even the four common components selected in Table~\ref{tab:mapping-table} are not widely specified across schedulers, and only Condor and Apollo specify them all.
A complete overview of all mapped stages is given in Tables \ref{tab:mapping-table-jobs} (job processing), \ref{tab:mapping-table-tasks} (task processing), \ref{tab:mapping-table-schedulers} (scheduler management), and \ref{tab:mapping-table-resources} (resource management).

We conclude that the schedulers mapped in this work are on average {\it underspecified}, and explore in Section~\ref{sec:experiments} the impact of underspecification on understanding the performance and the operation of schedulers. 
To refine our analysis, we focus next on the scheduler feature (column ``Feat.'' in Table~\ref{tab:mapping-table}).

\begin{figure}[!t]
\centering
\vspace*{-0.25cm}
\subfloat[Academia\label{fig:heatmap-components-academia}]{\includegraphics[width = 3.9cm]{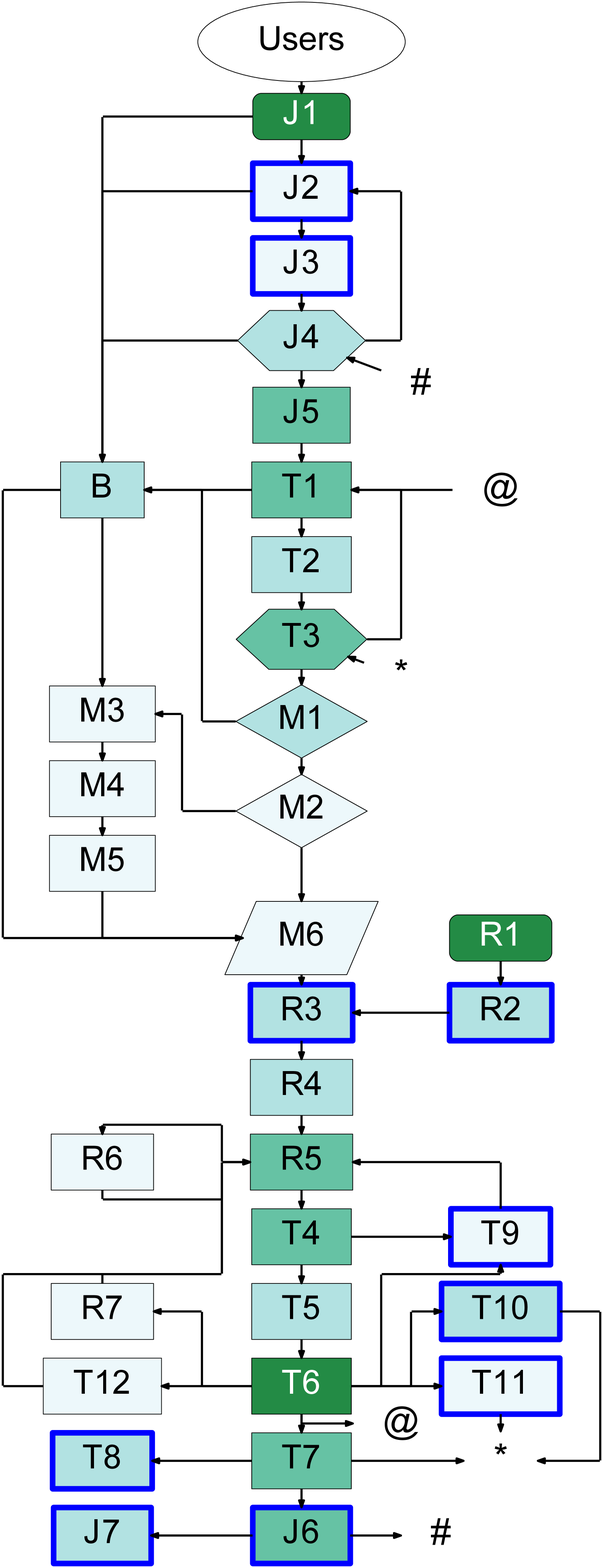}}
\subfloat[Industry\label{fig:heatmap-components-industry}]{\includegraphics[width = 4.6cm]{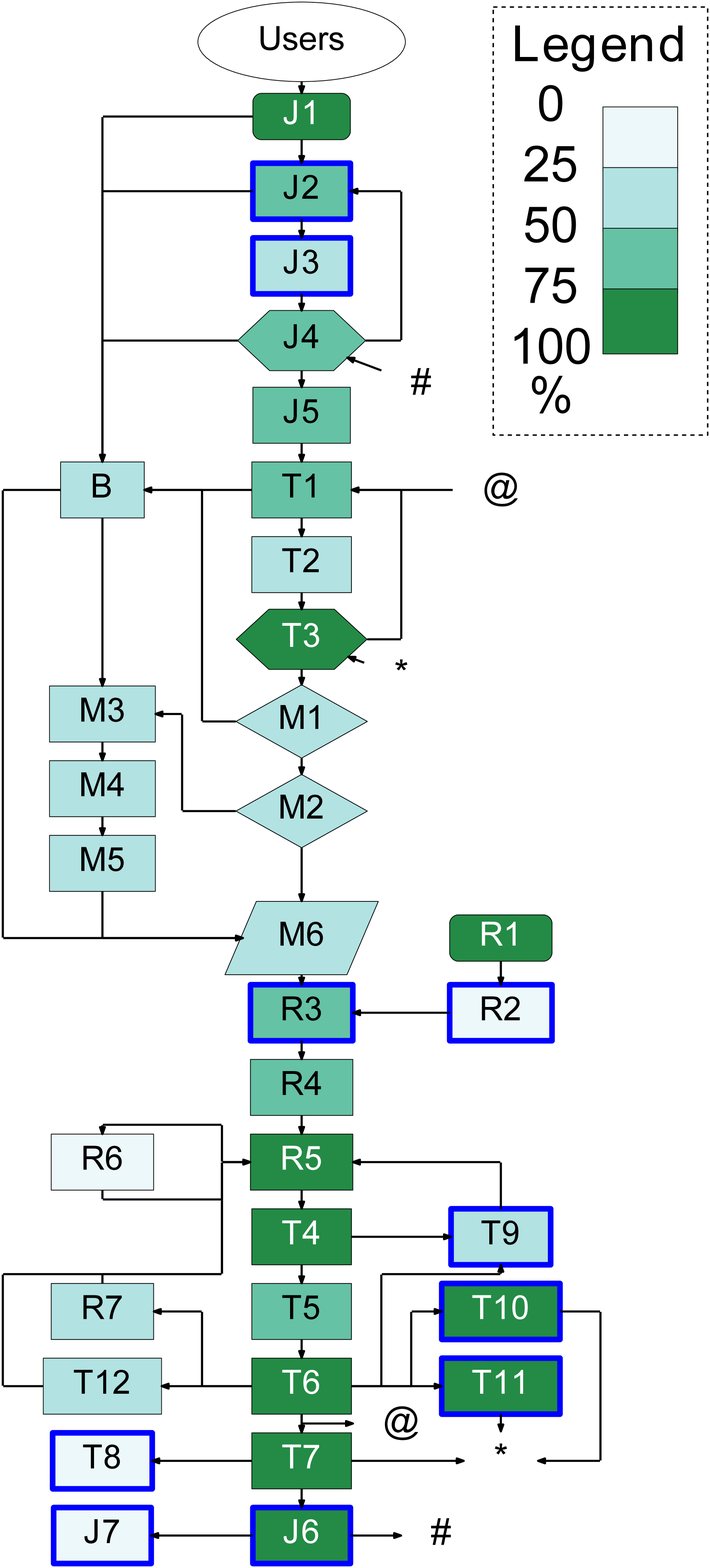}}
\vspace*{-0.25cm}
\caption{The specification of stages for academic and industrial schedulers. (Highlighted boxes relate to Table~\ref{tab:component-differences}. Intensities in the heatmap are depicted using a gradient, see Legend.)}
\label{fig:full-overview-academia-industry}
\end{figure}

\begin{table}
\centering
\caption{Full overview of job processing stages of schedulers mapped to the reference architecture.}
\label{tab:mapping-table-jobs}
\vspace*{-0.25cm}
\ra{1.2}
\begin{tabularx}{\columnwidth}{Xccccccc} \toprule
Scheduler & \texttt{J1} & \texttt{J2} & \texttt{J3} & \texttt{J4} & \texttt{J5} & \texttt{J6} & \texttt{J7} \\
\midrule
Condor & $\checkmark$ &  &  & $\sim$ & $\checkmark$ & $\checkmark$ & $\checkmark$ \\
Mesos & $\checkmark$ &  &  &  &  &  &  \\
Borg & $\checkmark$ & $\checkmark$ & $\sim$ & $\sim$ &  & $\checkmark$ &  \\
Fuxi & $\checkmark$ &  & $\sim$ & $\sim$ & $\checkmark$ & $\checkmark$ &  \\
Autopilot & $\checkmark$ &  &  & $\sim$ &  & $\checkmark$ &  \\
\midrule
Sparrow & $\checkmark$ &  &  & $\sim$ &  &  &  \\
Pegasus & $\checkmark$ & $\sim$ &  & $\sim$ & $\checkmark$ & $\checkmark$ & $\checkmark$ \\
Quincy & $\checkmark$ &  &  & $\sim$ & $\checkmark$ & $\sim$ &  \\
ICENI & $\checkmark$ &  & $\sim$ & $\sim$ & $\sim$ & $\checkmark$ &  \\
Firmament & $\checkmark$ &  &  &  &  &  &  \\
\midrule
Apollo & $\checkmark$ & $\checkmark$ & $\checkmark$ & $\checkmark$ & $\checkmark$ & $\checkmark$ &  \\
Askalon & $\checkmark$ &  &  & $\checkmark$ & $\checkmark$ & $\checkmark$ & $\checkmark$ \\
Triana & $\checkmark$ &  &  & $\sim$ & $\checkmark$ & $\sim$ &  \\
Dryad & $\checkmark$ & $\checkmark$ & $\sim$ & $\checkmark$ &  & $\checkmark$ &  \\
\bottomrule
\multicolumn{8}{l}{
Legend: $\checkmark$/$\sim$/space (\textvisiblespace{}) = full/partial/no match. }\\
\end{tabularx}
\end{table}

\begin{table*}
\centering
\caption{Full overview of task processing stages of schedulers mapped to the reference architecture.}
\label{tab:mapping-table-tasks}
\vspace*{-0.25cm}
\ra{1.2}
\begin{tabularx}{\linewidth-2.5in}{Xcccccccccccc} \toprule
Scheduler & \texttt{T1} & \texttt{T2} & \texttt{T3} & \texttt{T4} & \texttt{T5} & \texttt{T6} & \texttt{T7} & \texttt{T8} & \texttt{T9} & \texttt{T10} & \texttt{T11} & \texttt{T12} \\
\midrule
Condor & $\checkmark$ & $\checkmark$ & $\sim$ & $\checkmark$ & $\checkmark$ & $\checkmark$ & $\checkmark$ & $\checkmark$ & $\checkmark$ & $\checkmark$ &  & $\checkmark$ \\
Mesos &  &  & $\sim$ & $\checkmark$ &  & $\checkmark$ & $\checkmark$ & $\checkmark$ &  & $\checkmark$ &  &  \\
Borg & $\sim$ & $\sim$ & $\checkmark$ & $\checkmark$ & $\checkmark$ & $\checkmark$ & $\checkmark$ & $\checkmark$ &  & $\checkmark$ & $\checkmark$ & $\sim$ \\
Fuxi & $\sim$ &  & $\sim$ & $\sim$ &  & $\checkmark$ & $\checkmark$ &  &  & $\checkmark$ & $\checkmark$ & $\checkmark$ \\
Autopilot &  &  & $\sim$ & $\checkmark$ &  & $\checkmark$ & $\checkmark$ &  &  & $\checkmark$ & $\checkmark$ &  \\
\midrule
Sparrow &  & $\sim$ & $\sim$ & $\checkmark$ &  & $\checkmark$ & $\checkmark$ &  &  &  &  &  \\
Pegasus & $\checkmark$ &  & $\sim$ & $\checkmark$ & $\checkmark$ & $\checkmark$ & $\sim$ & $\checkmark$ &  & $\sim$ & $\checkmark$ &  \\
Quincy &  &  & $\sim$ & $\checkmark$ & $\checkmark$ & $\checkmark$ & $\sim$ &  & $\sim$ & $\checkmark$ &  & $\sim$ \\
ICENI & $\sim$ &  & $\sim$ &  & $\sim$ & $\checkmark$ & $\checkmark$ & $\sim$ &  &  &  &  \\
Firmament & $\sim$ & $\checkmark$ & $\checkmark$ & $\checkmark$ &  &  &  &  &  &  &  &  \\
\midrule
Apollo & $\checkmark$ & $\checkmark$ & $\checkmark$ & $\checkmark$ &  & $\checkmark$ & $\checkmark$ &  & $\checkmark$ & $\checkmark$ & $\checkmark$ &  \\
Askalon & $\sim$ &  & $\sim$ &  &  &  &  &  &  &  &  &  \\
Triana & $\checkmark$ & $\checkmark$ & $\sim$ & $\sim$ &  & $\checkmark$ & $\checkmark$ &  &  &  &  &  \\
Dryad & $\checkmark$ &  & $\checkmark$ & $\checkmark$ & $\checkmark$ & $\checkmark$ & $\checkmark$ &  & $\checkmark$ & $\sim$ & $\checkmark$ &  \\
\bottomrule
\multicolumn{13}{l}{
Legend: $\checkmark$/$\sim$/space (\textvisiblespace{}) = full/partial/no match. }\\
\end{tabularx}
\end{table*}

\begin{table}
\centering
\caption{Full overview of scheduler management stages of schedulers mapped to the reference architecture.}
\label{tab:mapping-table-schedulers}
\vspace*{-0.25cm}
\ra{1.2}
\begin{tabularx}{\columnwidth}{Xccccccc} \toprule
Scheduler & \texttt{B} & \texttt{M1} & \texttt{M2} & \texttt{M3} & \texttt{M4} & \texttt{M5} & \texttt{M6} \\
\midrule
Condor & $\checkmark$ & $\checkmark$ &  &  &  &  &  \\
Mesos &  &  &  &  &  &  &  \\
Borg &  &  &  &  &  &  &  \\
Fuxi &  &  &  &  &  &  &  \\
Autopilot &  &  &  &  &  &  &  \\
\midrule
Sparrow &  &  & $\checkmark$ & $\checkmark$ &  & $\checkmark$ & $\checkmark$ \\
Pegasus &  &  &  &  &  &  &  \\
Quincy &  &  &  &  &  &  &  \\
ICENI &  &  &  &  &  &  &  \\
Firmament & $\checkmark$ & $\checkmark$ &  &  &  &  &  \\
\midrule
Apollo & $\checkmark$ & $\checkmark$ & $\checkmark$ & $\checkmark$ & $\checkmark$ & $\checkmark$ & $\checkmark$ \\
Askalon & $\checkmark$ &  &  &  &  &  &  \\
Triana &  &  &  &  &  &  & $\sim$ \\
Dryad & $\checkmark$ & $\checkmark$ & $\checkmark$ & $\sim$ & $\sim$ & $\sim$ & $\sim$ \\
\bottomrule
\multicolumn{8}{l}{
Legend: $\checkmark$/$\sim$/space (\textvisiblespace{}) = full/partial/no match. }\\
\end{tabularx}
\end{table}

\begin{table}
\centering
\caption{Full overview of resource management stages of schedulers mapped to the reference architecture.}
\label{tab:mapping-table-resources}
\vspace*{-0.25cm}
\ra{1.2}
\begin{tabularx}{\columnwidth}{Xccccccc} \toprule
Scheduler & \texttt{R1} & \texttt{R2} & \texttt{R3} & \texttt{R4} & \texttt{R5} & \texttt{R6} & \texttt{R7} \\
\midrule
Condor & $\checkmark$ & $\checkmark$ & $\checkmark$ & $\checkmark$ & $\checkmark$ &  &  \\
Mesos & $\checkmark$ & $\checkmark$ &  & $\checkmark$ & $\checkmark$ &  & $\checkmark$ \\
Borg & $\checkmark$ &  & $\checkmark$ & $\checkmark$ & $\checkmark$ &  & $\sim$ \\
Fuxi & $\checkmark$ &  &  & $\checkmark$ & $\sim$ &  & $\checkmark$ \\
Autopilot & $\sim$ &  & $\checkmark$ &  & $\sim$ &  &  \\
\midrule
Sparrow &  &  &  &  &  &  &  \\
Pegasus & $\checkmark$ &  &  &  & $\sim$ & $\checkmark$ &  \\
Quincy & $\checkmark$ &  &  &  & $\checkmark$ &  &  \\
ICENI & $\checkmark$ & $\checkmark$ & $\checkmark$ &  & $\checkmark$ &  &  \\
Firmament & $\checkmark$ &  & $\checkmark$ & $\checkmark$ & $\checkmark$ &  &  \\
\midrule
Apollo & $\checkmark$ &  & $\checkmark$ & $\checkmark$ & $\checkmark$ &  &  \\
Askalon & $\checkmark$ &  &  &  &  &  &  \\
Triana & $\sim$ &  &  &  & $\sim$ &  &  \\
Dryad & $\checkmark$ &  & $\checkmark$ & $\sim$ & $\sim$ &  &  \\
\bottomrule
\multicolumn{8}{l}{
Legend: $\checkmark$/$\sim$/space (\textvisiblespace{}) = full/partial/no match. }\\
\end{tabularx}
\end{table}
    
\subsection{Academia- vs. Industry-Designed Schedulers} 
\label{sec:validation:industry-academia}

To understand the differences between schedulers designed by academia and by industry,
we repeat the mapping process used in the previous section, independently, for the 8 academic and 6 industrial schedulers considered in this work (labeled ``A'' and ``I'', respectively, in column ``Feat.'', in Table~\ref{tab:mapping-table}). 

Figures~\ref{fig:heatmap-components-academia} and~\ref{fig:heatmap-components-industry} depict aggregated results for academia and industry-designed schedulers, respectively. 
Color-gradients depict the aggregated values, using the method used earlier for the group-aggregates included in Table~\ref{tab:mapping-table}.
The two figures indicate there are some stages where academia and industry underspecify similarly, but many more stages where academia and industry diverge.
For example, \texttt{T1} is defined by [50\%,75\%) of both academic and industry schedulers, whereas \texttt{R5} is defined by [50\%,75\%) of academic schedulers and [75\%,100\%] (more) of the industry schedulers. 
Similarly, \texttt{R6} is rarely specified for all schedulers, indicating resource provisioning (linked to autoscaling, \texttt{R7}) remains understudied.

To quantify differences in underspecification, Table~\ref{tab:component-differences} summarizes the 10 components having the largest absolute difference in specification ({\it match}), between academia and industry.
We observe that task preemption~(\texttt{T10}) and recovery~(\texttt{T11}) are often specified in work describing industrial systems, yet are underspecified by academia. 
The industry preference for such features, which have no generic techniques and are thus perceived as engineering-heavy in the academia, can be explained through a simile. 
An organization needing to select a Linux operating system will need to choose between performance and stability.
The organization could choose to take on Linux distributions with higher technical risk, and deploy Fedora or Debian Unstable. These distributions focus on quick and innovative releases, which may improve performance, but may also break more often.
Alternatively, the organization could choose for lower risk, installing therefore 
Red Hat Enterprise Linux (or CentOS), Debian, or an Ubuntu LTS release. They have well-tested, well-engineered features.

Specifying the list of eligible jobs~(\texttt{J2}) is also done more often by industry, relatively to the academia.
Conversely, filtering based on authorization~(\texttt{R2}) and job cleanup~(\texttt{J7}) are stages underspecified more by industry than by academia.
An alternation of underspecification can occur for stages that from a chain of processing operations,
for example, job completion~(\texttt{J6}, underspecified less by industry) and job cleanup~(\texttt{J7}, underspecified less by academia).

There could be several reasons for these differences in reporting.
Correctness of computation~({\tt T10}) and the consistent state of the system ({\tt T11}) are perhaps more important for industry, as they have to adhere to (often strict) SLAs, whereas in the academia these aspects may be seen as either a separate concern (thus, they will be mentioned in other, specialized articles) or a technical issue (thus, they will be less specified in peer-reviewed publications).
Authorization issues~(\texttt{R2}) may not be seen as important when jobs run in a private datacenter or trusted cloud; if this is the case, stage~{\tt R2} should become more important for industry-designed schedulers, proportionally with the adoption of public-cloud services.
Another reason can be practical experience, e.g., job cleanup~(\texttt{J7}) must always be implemented following (or as part of) job completion~(\texttt{J6}); to a lesser extent, we observe this situation also for tasks (\texttt{T7} and, notably, \texttt{T8}).

\subsection{Pre-2010 Schedulers vs. Schedulers Since 2010}
\label{sec:validation:publication}

\begin{figure}[!t]
\centering
\subfloat[Published before 2010\label{fig:heatmap-components-before2010}]{\includegraphics[width = 3.9cm]{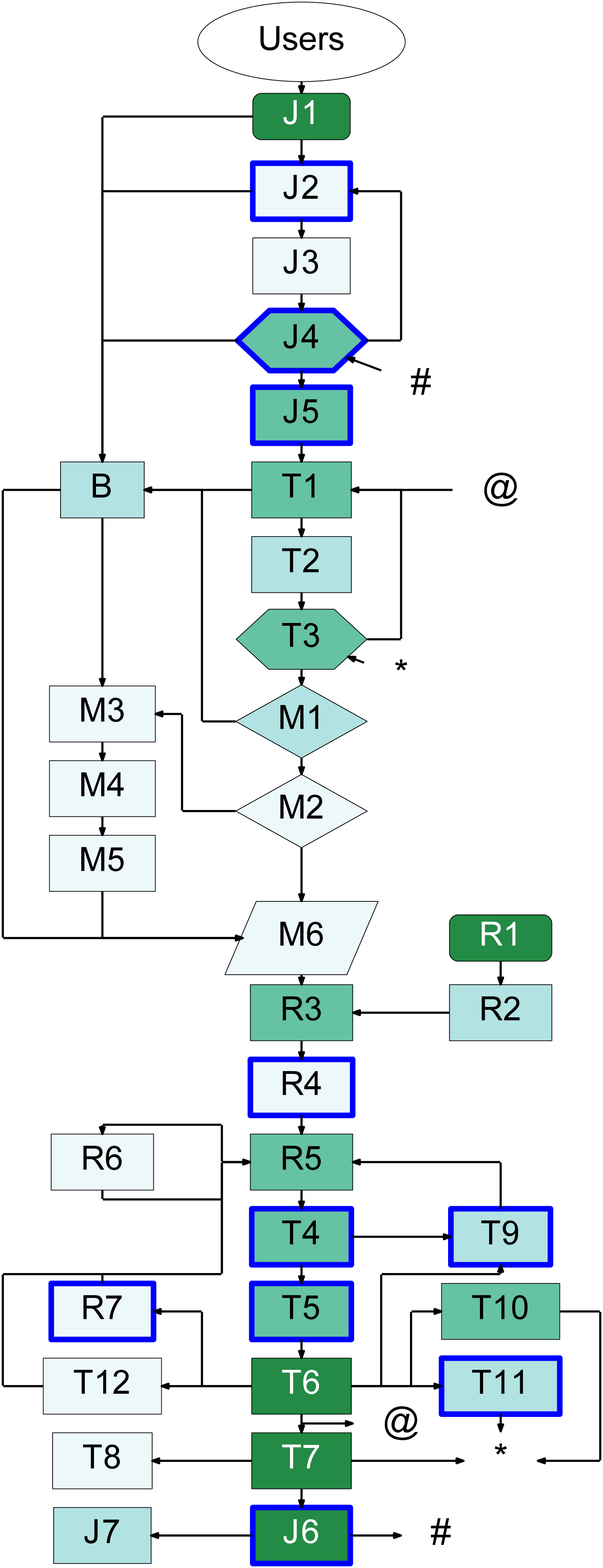}}
\subfloat[Published in/after 2010\label{fig:heatmap-components-after2010}]{\includegraphics[width = 4.6cm]{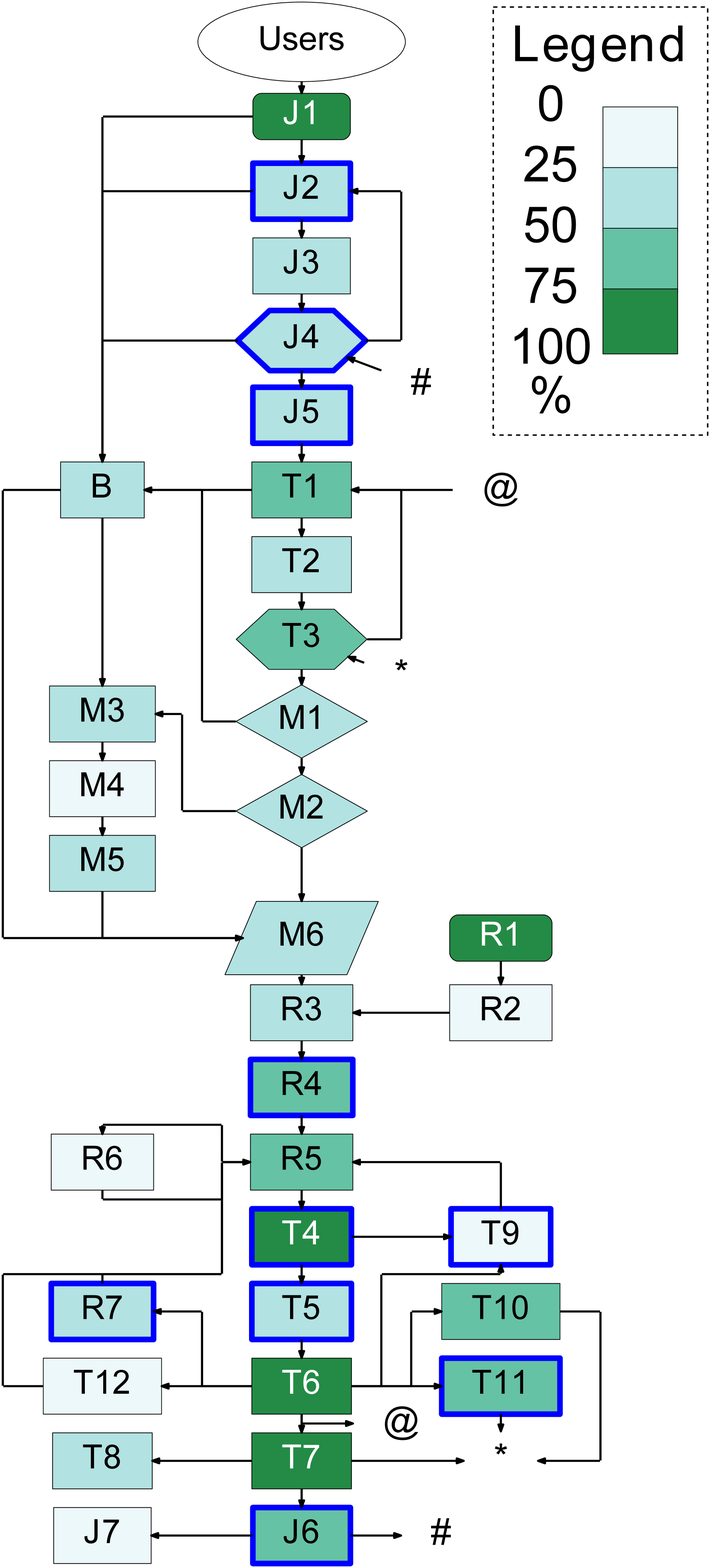}}
\vspace*{-0.25cm}
\caption{The specification of stages for for old, pre-2010 and newer schedulers. (Highlighted boxes relate to Table~\ref{tab:age-component-differences}. Intensities in the heatmap are depicted using a gradient, see Legend.)}
\label{fig:full-overview-publication}
\vspace*{-0.5cm}
\end{figure}

\begin{table}[!t]
	\centering
	\caption{The top-10 stages with the largest difference in match (\%), for old, pre-2010 (O) vs. newer (N) schedulers.}
	\label{tab:age-component-differences}
    \vspace*{-0.25cm}
	\begin{tabularx}{\columnwidth}{lrrrrrrrrrr}
		\toprule
		Publish     & \texttt{R4}   & \texttt{R7}   & \texttt{T4}   & \texttt{T11}  & \texttt{J6}    & \texttt{J5}   & \texttt{J4}   & \texttt{T9} & \texttt{T5} & \texttt{J2}   \\ \midrule
        O  & 21 & 0    & 64 & 29 & 86  & 64 & 64 & 36 & 50 & 14 \\
		N & 71 & 36 & 93 & 57 & 57  & 43 & 43 & 14 & 29 & 35  \\
		 \bottomrule
	\end{tabularx}
\end{table}

To understand if there has been a change of focus on scheduling-stages in recent designs, we repeat the mapping process
used in Section~\ref{sec:validation:industry-academia}, and map in turn the schedulers designed prior to 2010 and since 2010. 

Similarly to the previous section, 
we visualize the aggregate results in Figure~\ref{fig:full-overview-publication} and summarize the largest ten differences in Table~\ref{tab:age-component-differences}.
From this table, we note a sharp increase in {\tt R4}, {\tt R7}, {\tt T4}, and \texttt{T11} for schedulers introduced since 2010.
We believe this correlates with the increase of dynamicity in real-world environments, leading to increased focus on dynamic resource provisioning and allocation.
In contrast, the components {\tt J4-6}, {\tt T9}, and \texttt{T5} show a drop of focus in recent years. This is expected: job completion ({\tt J6}) and cleanup ({\tt J5}), and task setup ({\tt T5}) have used similar techniques for more than a decade, and thus have become tacit knowledge.
Interestingly, \texttt{T9} (task replication) has also been less featured, which could be explained by both standardization and by the realization that the resource waste incurred by large-factor replication is too expensive.

\section{Experiments with the Reference Architecture}
\label{sec:experiments}

In this section, our goal is to understand the impact of stage-underspecification (defined in Section~\ref{sec:validation}) on the performance and operation of real-world scheduling approaches.
This gives evidence of the usefulness of the reference architecture ({\it R2}). 

To achieve our goal, we conduct an experimental evaluation of scheduling approaches, using {\it dynamic analysis}, that is, analysis based on relevant input traces considered by the scheduling system at runtime. 
We conduct simulation experiments to evaluate running long-term workloads and real-world experiments to evaluate only the algorithmic (policy) part of the scheduler.

\subsection{Experiment Setup} \label{sec:experiments:setup}\label{sec:exp:setup}
 
\subsubsection{Overview} Table~\ref{tab:experiment-overview} summarizes the experimental setup. 
Experiments use an implementation of the reference architecture, in simulation for the experiment in Section~\ref{sec:experiments:underspecification}, and in a real-world environment for the experiment in Section~\ref{sec:experiments:complexity}. The experiments use as input relevant traces collected from real-world environments, and shared by the community through the Grid Workloads Archive~\cite{DBLP:journals/fgcs/IosupLJADWE08}. 
The experiments use a diverse set of metrics to quantify the performance of scheduler workload-management and operation.

\subsubsection{Simulation and Real-World Experiments} 
For simulations, we extend the community-driven, open-source OpenDC simulation platform~\cite{opendc}. 
The simulated setup is typical of event-driven simulation of (federated) clusters, and has been implemented similarly by instruments such as CloudSim, GroudSim, etc.
We prototype the reference architecture, making configurable the stages \texttt{T1-2} and \texttt{R4-5}. 
These stages are constructed based on policies used by real-world schedulers.

For real-world experiments, we run only the abstract parts of the architecture (the algorithms) on a real-world machine, configured as described in Section~\ref{sec:exp:setup}.5. 
We subject the algorithms to the incoming load also used during simulations, but for the real-world experiments the scheduler does not follow through with the enforcement of the policy-decisions.
This allows the real-world experiments to complete without consuming real datacenter resources, yet to reveal the dynamic behavior of the algorithms under realistic load (in contrast to static, worst-case analysis).

\subsubsection{One experiment} All setups are repeated 32 times, preceded by 4 (discarded) warm-up iterations, to reduce the influence of the underlying platform and cold caches on the measurements. 
Results, which include hundreds to thousands of samples due to the large input workloads (see {\it Workloads}), are processed statistically
and verified independently.

\subsubsection{Workloads} We experiment here only with workloads of workflows, which are the most complex type of workload supported by the general model. Experimenting with other models, e.g., bags of tasks, while useful, would not have used the full features of workflow-capable schedulers.

We use two traces collected from real-world datacenter-like environments by the community, \wlaskalon~\cite{DBLP:journals/fgcs/IosupLJADWE08} and \wlchronos~\cite{DBLP:conf/icac/MaISI17}. 
\wlaskalon{} is an engineering workload from a grid cluster that uses workflows to simulate chemical processes.
\wlchronos{} is an industrial workload from a private cloud that uses workflows to process data collected from an IoT production-environment monitoring industrial equipment.
Both workloads have been collected independently from this study, and feature in previous work of the community.
 Table~\ref{tab:exp:setup:wl} summarizes the characteristics of both workloads: \wlaskalon{} has more complex workflows, 
\wlchronos{} includes a large job-burst at start.

\subsubsection{Datacenter topology} 
We consider in this work a datacenter with 32 common-off-the-shelf resources (machines). Half (16) of these machines contain each an Intel i7 (v6) processor, with 4 cores, at a clock rate of 4.1GHz.
The other half contain each an Intel i5 (v6) processor, with 2 cores, at a clock rate of 3.5GHz.

For real-world experiments,
the scheduler 
runs
on
the
Intel i7 processor, and Java 8 (1.8.0\_162, Oracle engine).

\subsubsection{Metrics} our analysis uses four traditional metrics:
(1) Task response time (\emph{TRT}): time elapsed from task submission to task completion, 
(2) Job makespan (\emph{JMS}): time elapsed from the first task-submission of a job, to the last completion of a task in the job,
(3) Normalized Job-Schedule Length~\cite{kwok1998benchmarking} (\emph{NJSL}): job makespan normalized by the length of the critical path (the shortest possible execution time of the job),
(4) Job waiting time (\emph{JWT}): time elapsed from the first task-submission of a job, to the first start of a task of that job.
We also use, for real-world experiments, the metric: 
(5) Scheduler-stage time-complexity (\emph{TC}):  the runtime, or time elapsed since start and until completion, for the stage. 

\begin{table}[!t]
	\centering
	\caption{Experiment configurations.}
	\label{tab:experiment-overview}
    \vspace*{-0.15cm}
	\begin{tabularx}{\columnwidth}{lXccc}
		\toprule
		 Sec. & Experiment focus & Workload & Metrics   \\ \midrule
		\S\ref{sec:experiments:underspecification} & Underspecification & All (\S\ref{sec:exp:setup}.4)  & TRT, JMS, NJSL, JWT \\
		\S\ref{sec:experiments:complexity}  & Time complexity &  All (\S\ref{sec:exp:setup}.4) & TC  \\ \bottomrule
	\end{tabularx}
    \vspace*{-0.35cm}
\end{table}

\begin{table}[b]
\vspace*{-0.35cm}
	\centering
	\caption{Workload characteristics.}
	\label{tab:exp:setup:wl}
    \vspace*{-0.15cm}
	\begin{tabularx}{\columnwidth}{llXcc}
		\toprule
		 Workload & Env. & Application domain & Workflows & Tasks   \\ \midrule
		\wlaskalon & Grid & Engineering, chemistry &  757 & 45,786 \\
		\wlchronos & Cloud & Industrial, IoT        & 1,024  & 3,072  \\ \bottomrule
	\end{tabularx}
\end{table}

\subsection{Impact of Underspecification on Performance (Simulation)} \label{sec:experiments:underspecification}

We investigate the impact of scheduler underspecification on performance. 
We focus on components that must exist (that is,  based on their own publications, the correct operation of the schedulers under study requires these components), yet have been left underspecified by their authors.
Although different policies have been shown in the past to lead to different performance,  
ours is the first study to show the impact of {\it underspecified} policies across different components of the {\it same scheduler}.
Such underspecification can hamper the reproducibility of a proposed scheduling system.
To illustrate this, we use the Borg publication~\cite{Verma2015} as an example. 
The publication does not specify its task-sorting, task-allocation, and others policies, although it requires their presence.
To highlight the importance of specification, {\it in this experiment we simulate a Borg-like scheduler, equipped with different policies likely to be used in practice}. 

\begin{figure}[!t]
\subfloat[TRT distribution per task\label{fig:underspecification:eng:response-time-plot}]{\includegraphics[width=\linewidth]{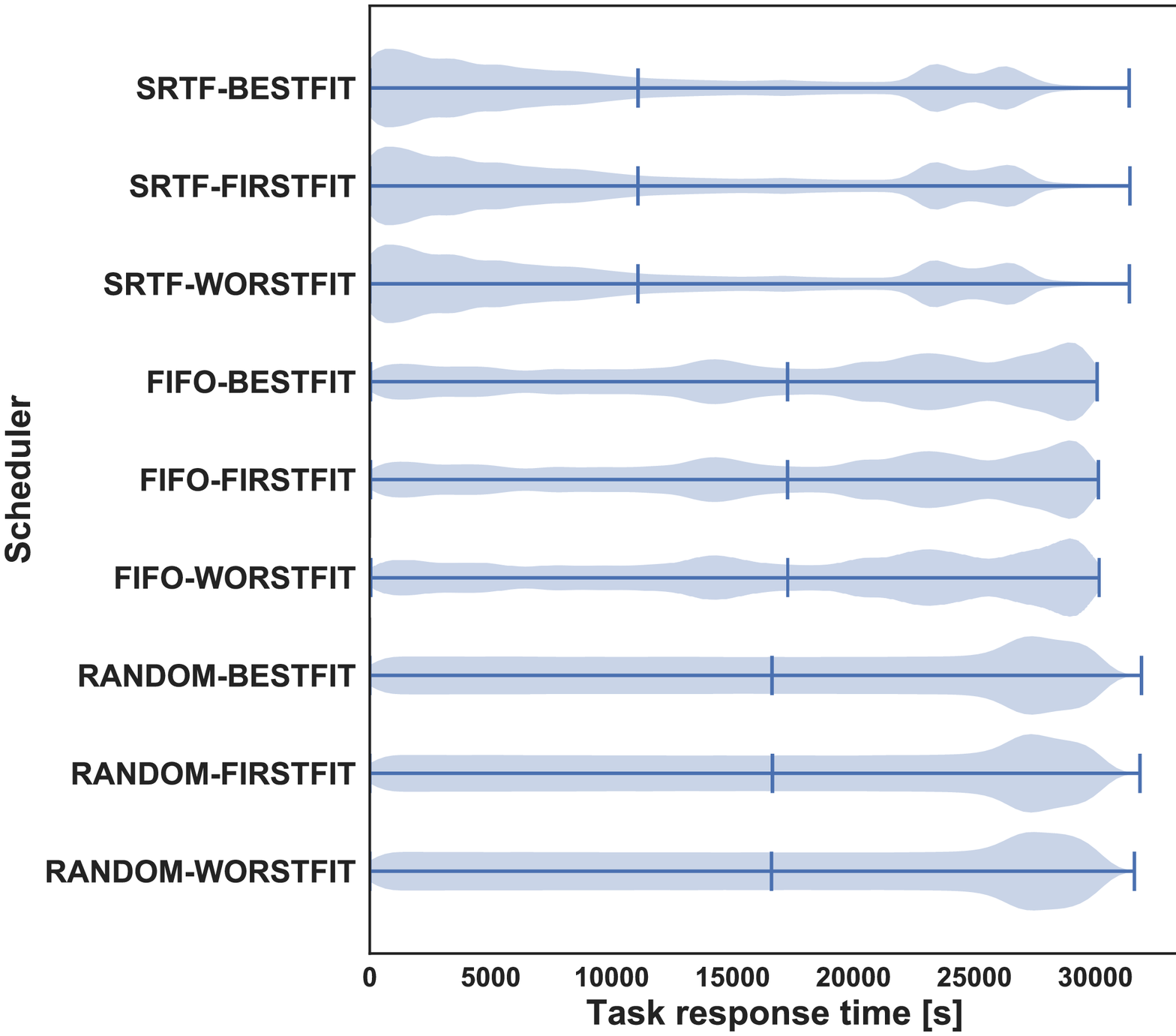}}\\
\subfloat[JMS distribution per job\label{fig:underspecification:eng:makespan-plot}]{\includegraphics[width=\linewidth]{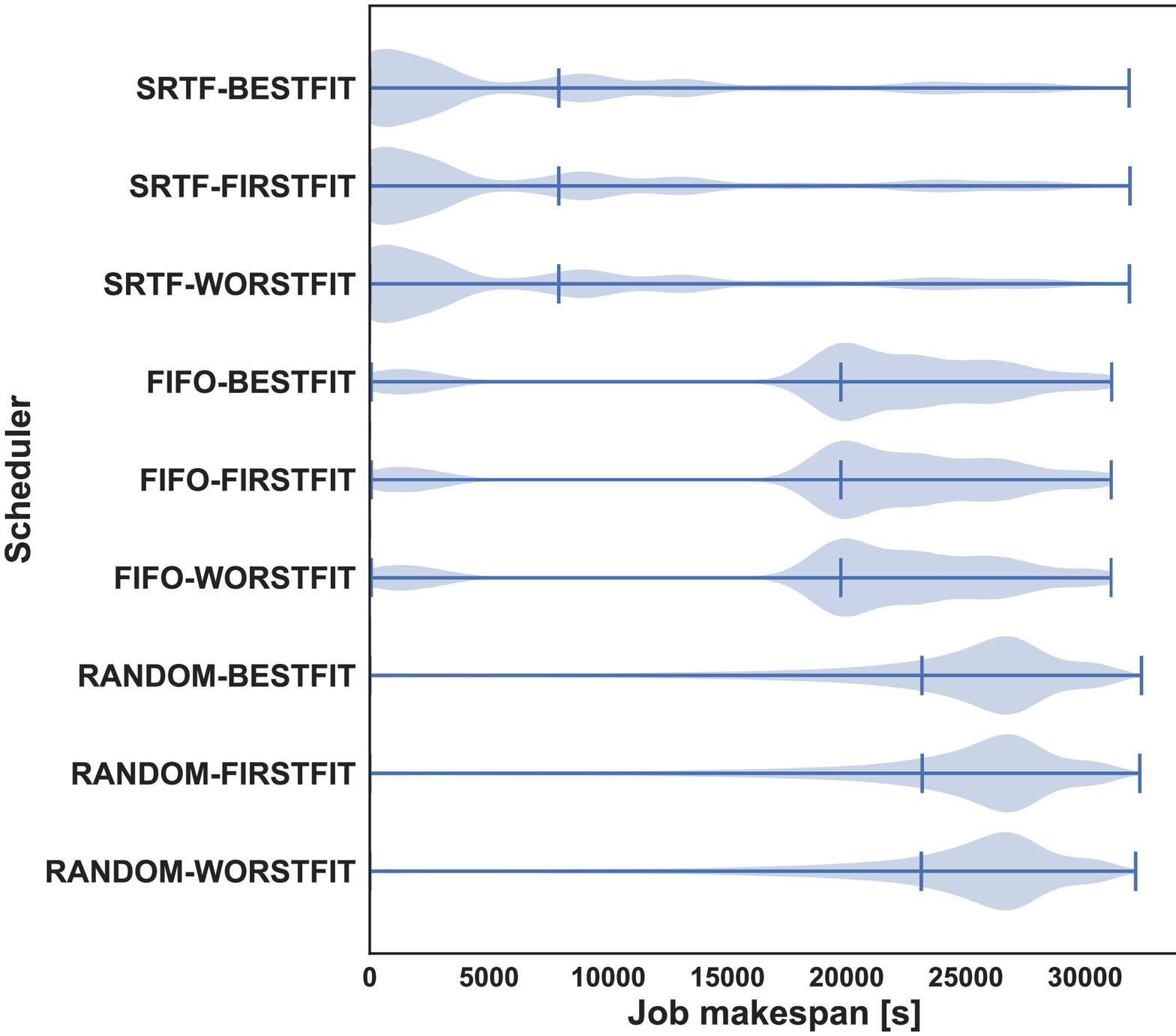}}
    \caption{Performance per scheduler configuration. The vertical bar in the middle indicates the average value for (a) response time, (b) job makespan. (Workload: \wlaskalon{}.)}
    \label{fig:underspecification:eng}
    \vspace*{-0.35cm}
\end{figure}

\begin{figure}[!t]
\subfloat[TRT distribution per task\label{fig:underspecification:ind:response-time-plot}]{\includegraphics[width=\linewidth]{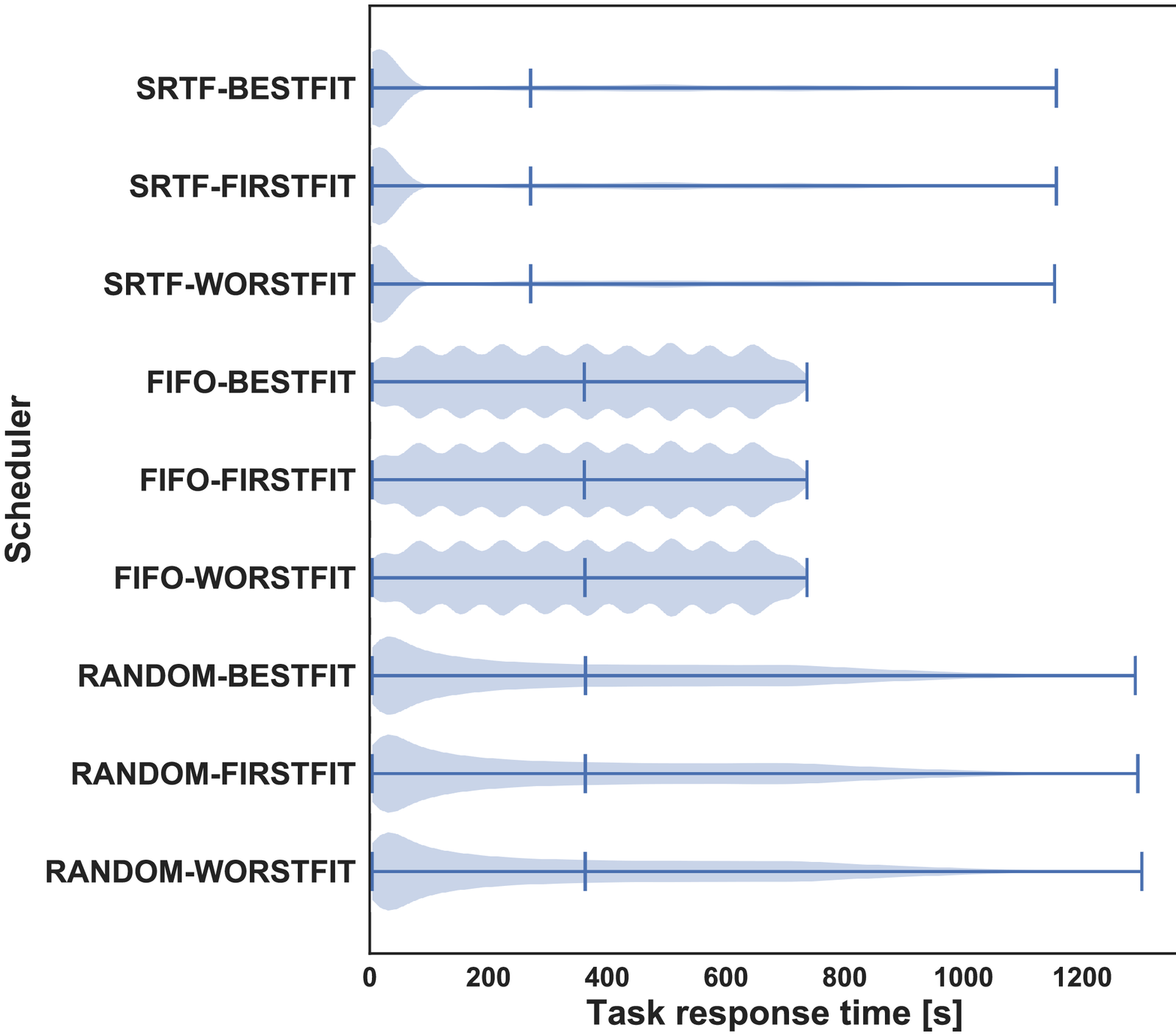}}\\
\subfloat[JMS distribution per job\label{fig:underspecification:ind:makespan-plot}]{\includegraphics[width=\linewidth]{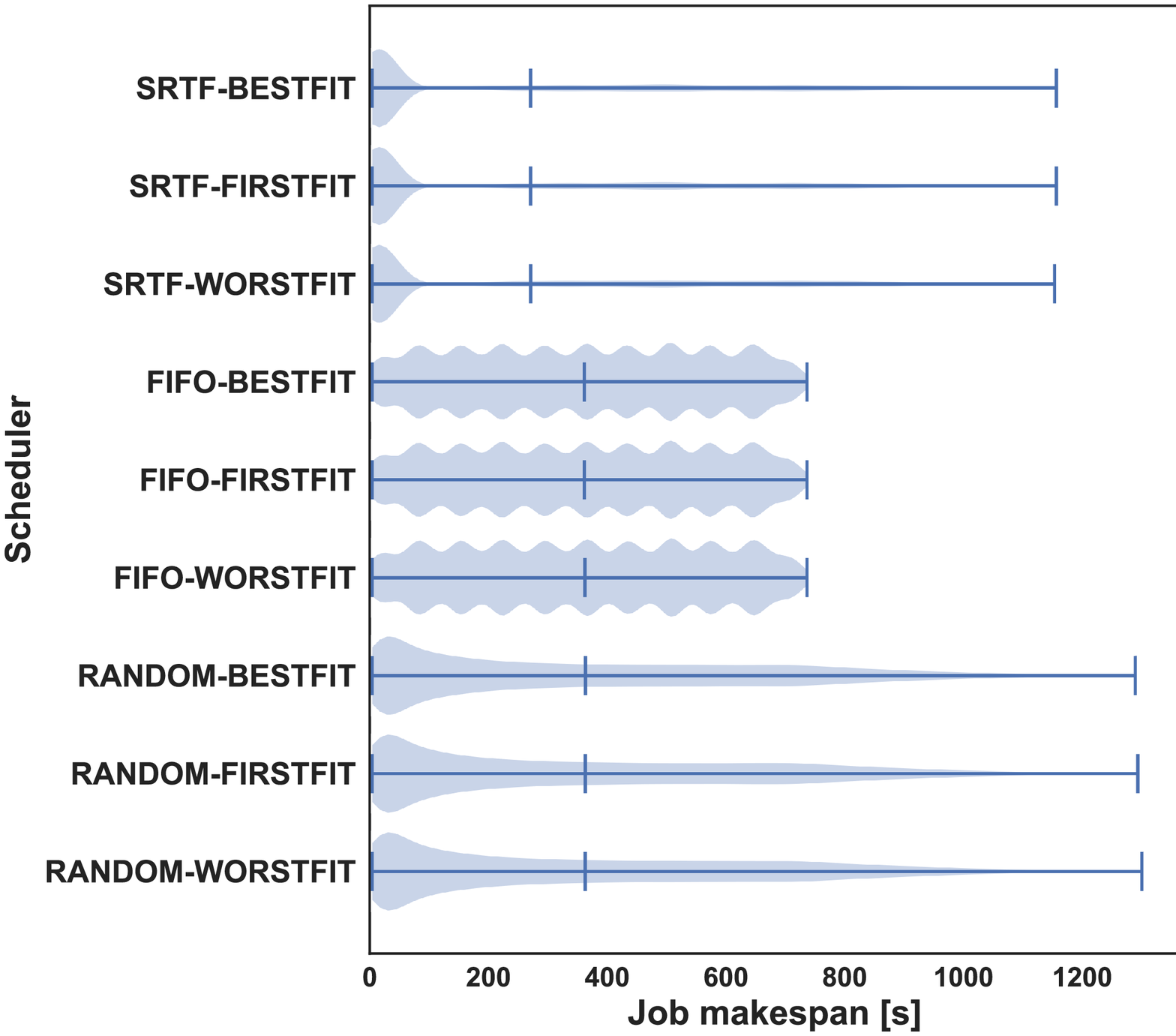}}
    \caption{Performance per scheduler configuration. The vertical bar in the middle indicates the average value for (a) response time, (b) job makespan. (Workload: \wlchronos{}.)}
    \label{fig:underspecification:ind}
    \vspace*{-0.35cm}
\end{figure}

Figures~\ref{fig:underspecification:eng:response-time-plot} and \ref{fig:underspecification:ind:response-time-plot} depict the TRT distribution measured for Borg, when equipped with different sorting and allocation policies. 

Overall, the choice of task-sorting policy (\texttt{RANDOM}, \texttt{FIFO}, or \texttt{SRTF}) is correlated with the distribution of task turnaround times; the allocation policy shows less distinctive results.
\texttt{SRTF} gives the best average TRT. 

We conduct a similar analysis, per job. Figures~\ref{fig:underspecification:eng:makespan-plot} and \ref{fig:underspecification:ind:makespan-plot} depict the results.
The distribution of JMS is more clearly distributed around a limited number of peaks than the distribution of TRT.
\texttt{SRTF-BESTFIT} performs significantly better compared to the other two configurations.
However, unlike for TRT, for JMS the \texttt{FIFO} policy gives a better average than \texttt{RANDOM}.

Tables~\ref{tab:underspecification:eng:metrics} and \ref{tab:underspecification:ind:metrics} extend the performance analysis with the average NJSL and the average JWT.
As for JMS, \texttt{SRTF-BESTFIT} again emerges as best-performing for the NJSL metric. 
The policy of running the shortest remaining-time first reduces the slowdown, by design.

For JWT, the \texttt{FIFO-BESTFIT} configuration performs best 
because the \texttt{FIFO} policy for task-sorting preserves the order of submission, which leads to tasks being served with less time spent waiting, on average.

We conclude that {\it the underspecification of scheduler-stages leads to statistically different performance results}.
Even for experts, the results necessarily depend on the configuration (interpretation) of policies for scheduler-stages.

This observation, and the underspecification exemplified in Table~\ref{tab:mapping-table}, provide strong motivation for more precise specifications and details of schedulers in scientific literature---this is important in practice, and a new aspect of reproducibility~\cite{DBLP:conf/sc/HoeflerB15}.

\begin{table}[!t]
	\centering
	\caption{Average job makespan (JMS), normalized schedule length (NSL), and job waiting time (JWT) per scheduler (Workload: \wlaskalon{}).}
	\label{tab:underspecification:eng:metrics}
    \vspace*{-0.15cm}
	\begin{tabularx}{\columnwidth}{Xrrr}
    \toprule
Scheduler & Avg. JMS [s] & Avg. NJSL & Avg. JWT [s]   \\ \midrule
SRTF-BESTFIT & 7,929 & 5 & 3,134 \\
SRTF-FIRSTFIT & 7,927 & 5 & 3,134 \\
SRTF-WORSTFIT & 7,927 & 5 & 3,135 \\
FIFO-BESTFIT & 19,751 & 32 & 2,478 \\
FIFO-FIRSTFIT & 19,751 & 32 & 2,480 \\
FIFO-WORSTFIT & 19,748 & 32 & 2,478 \\
RANDOM-BESTFIT & 23,156 & 206 & 4,789 \\
RANDOM-FIRSTFIT & 23,171 & 197 & 4,808 \\
RANDOM-WORSTFIT & 23,132 & 196 & 4,815 \\
        \bottomrule
	\end{tabularx}
\end{table}

\begin{table}[!t]
	\centering
	\caption{Average job makespan (JMS), normalized schedule length (NSL), and job waiting time (JWT) per scheduler (Workload: \wlchronos{}).}
	\label{tab:underspecification:ind:metrics}
    \vspace*{-0.15cm}
	\begin{tabularx}{\columnwidth}{Xrrr}
    \toprule
Scheduler & Avg. JMS [s] & Avg. NJSL & Avg. JWT [s]   \\ \midrule
SRTF-BESTFIT & 270 & 25 & 262 \\
SRTF-FIRSTFIT & 270 & 25 & 262 \\
SRTF-WORSTFIT & 270 & 25 & 262 \\
FIFO-BESTFIT & 361 & 50 & 353 \\
FIFO-FIRSTFIT & 361 & 50 & 353 \\
FIFO-WORSTFIT & 362 & 50 & 354 \\
RANDOM-BESTFIT & 363 & 51 & 355 \\
RANDOM-FIRSTFIT & 363 & 51 & 355 \\
RANDOM-WORSTFIT & 363 & 51 & 355 \\
        \bottomrule
	\end{tabularx}
\end{table}

\begin{figure}[!t]
	\includegraphics[width=\linewidth]{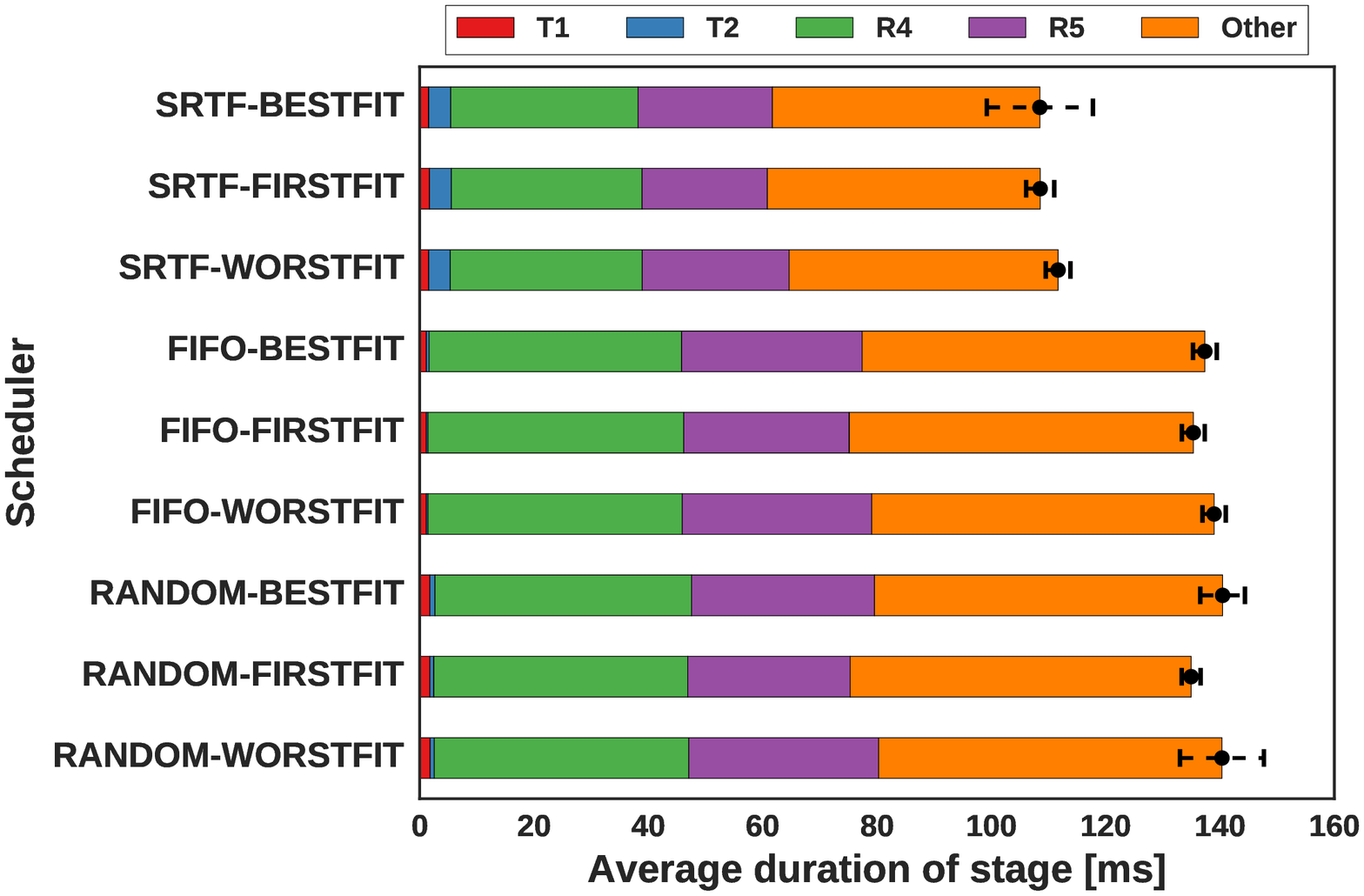}
    \vspace*{-0.5cm}
	\caption{Time complexity decomposition by stage, per scheduler configuration. (Workload: \wlchronos{}.)}
	\label{fig:complexity-plot-stacked:ind}
    \vspace*{-0.25cm}
\end{figure}

\begin{figure}[!t]
	\includegraphics[width=\linewidth]{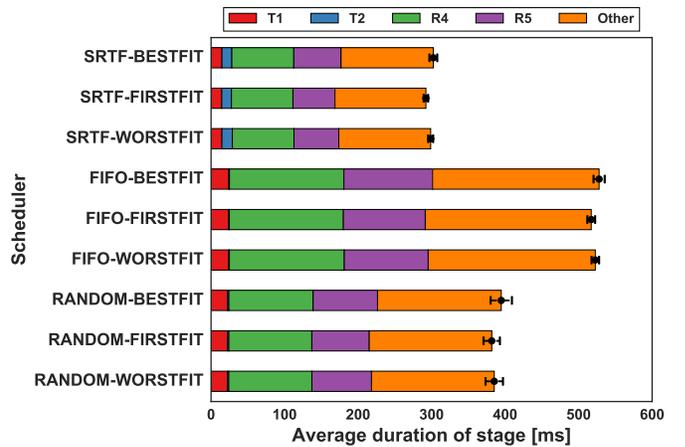}
    \vspace*{-0.5cm}
	\caption{Time complexity decomposition by stage, per scheduler configuration. (Workload: \wlaskalon{}.)}
	\label{fig:complexity-plot-stacked:eng}
    \vspace*{-0.25cm}
\end{figure}

\subsection{Stage-Complexity at Runtime (Real-World Experiment)} \label{sec:experiments:complexity}

We investigate the impact of underspecification of scheduler-stages on the computational complexity (time) of the scheduling process. Unlike traditional, worst-case, asymptotic analysis, we use in this work dynamic analysis: {\it we implement the scheduling algorithms and run them in real-world experiments, with streaming incoming workload}.
Similarly to Section~\ref{sec:experiments:underspecification}, we explore
three policies each for task-sorting ({\tt T2}) and 
for resource-allocation ({\tt R5}), and measure in particular the performance of stages {\tt T1-2} and {\tt R4-5}.

Figures~\ref{fig:complexity-plot-stacked:eng} and \ref{fig:complexity-plot-stacked:ind} depict the time-complexity decomposition, by stage,
for nine scheduler-configurations.
The \texttt{SRTF} task-sorting policy, which delivers the lowest average JMS and NJSL (Section~\ref{sec:experiments:underspecification}), leads to a significant increase in the time-complexity of the \texttt{T2} stage, compared to the other task-sorting policies.
However, the average overall time complexity of schedulers using {\tt SRTF} is still the lowest in our experiments, because it produces higher task-throughput and lower TRT (see also Figures~\ref{fig:underspecification:eng:response-time-plot} and \ref{fig:underspecification:ind:response-time-plot}); in turn, this reduces the task-queue length, and thus the durations of stages that traverse it, in particular, {\tt R4-5}.

Similarly to Section~\ref{sec:experiments:underspecification}, we conclude from these experiments that \emph{the stage-policy can have a significant, non-trivial impact on both single-stage duration and the overall duration of the scheduling iteration.} The non-trivial impact is derived from the {\it dynamic interplay} between the task throughput, which is likely improved by more compute-intensive policies, and the runtime to get a decision from the scheduler, which is improved by using less compute-intensive policies. The complexity of the scheduling pipeline, which can include policies for each stage in our reference architecture, can further increase the dynamic interplay. 

{\it The underspecification of stage-policies may be costly for practitioners}, because, 
when trying to re-use best-practices, or a de facto standard, or a new policy just created elsewhere, underspecification leads to interpretation and guess-work, which in turn can lead to implementing a different scheduler than the original authors did actually evaluate.
Similarly, when practitioners create their own alternatives, underspecification could hamper the work of operators and/or QA teams.

\section{Threats to Validity} \label{sec:threats}

We cover in this section the threats to the validity of our work, and our steps to mitigate them.

\subsection{Completeness and Overspecification of the Reference Architecture}\label{sec:threats:completeness}

As threats, completeness, that is, that our reference architecture does not cover even the necessary parts, and overspecification, that is, that our reference architecture covers much more than the strictly necessary components, are important.

We perceive the {\it completeness} of the reference architecture as the biggest threat to the validity of this work.
We took several steps to mitigate this risk. 
First, the mapping process was conducted by two reviewers, working independently, to facilitate the identification of issues with the definition and structure of the reference architecture, to promote discussion of components from independent points of view, and to explore a process to reconcile if a disagreement occurred.
Second, the fourteen systems that we have mapped onto the reference architecture cover the dimensions of background (academia and industry), age (pre and post 2010), deployment (single- vs. multi-cluster), and support for complex jobs (merely single complex jobs, bags of tasks, workflows).
Third and last, the selected systems have been analyzed and presented to the systems community and are some of the highest cited or reportedly used in practice, showing maturity and perceived usefulness.

Another item that could pose a threat to validity is the {\it underspecification analysis},  where certain schedulers may consciously omit certain components as they introduce, e.g., additional operational risk, or are not required. 
Although indeed the omission of specific components could be deliberate, in our heatmap analysis we focus on what the community itself deems important in reporting in their submitted work---in other words, if at least a large part of the community describes a component, that component must be important enough to warrant the attention of all experts in the field.
This allows us to compare the focus of different sides of the community, e.g., academia versus industry.
Moreover, to avoid experimenting with uncommon components, in our validation experiments (Section~\ref{sec:experiments}) we explicitly focus on components that \textit{must} be present in every scheduler for it to function, yet are not specified.
Finally, as a reference architecture by its nature is a super-set of components, not all systems will match all components.

\subsection{Validity of the Reference Architecture and Simulator}\label{sec:threats:validity}

The validity of any model and of the simulator enacting it are traditional threats to the validity of work based on such constructs.

The use of a {\it simulator}, instead of a real-world setup, could be another threat to validity of our experimental results.
To address this threat, we have validated our simulator (i) manually, through real-world workloads and verifying the outcomes, (ii) automatically, by running known the combination of (workload, environment) of prior work and verifying the outcome matches the report of prior work. However, the use of the currently existing alternatives, that is, real-world experimentation and mathematical analysis (e.g., hierarchical and queuing models), could suffer from the same and even deeper problems---real-world experiments is unlikely to use long-term traces or large-scale environments, and analysis could suffer from either terseness or over-fitting.

{\it Omitting I/O and network} from the reference architecture could pose another threat to the validity of our work.
However, the allocation policy that is employed in the scheduler is responsible for this. 
While we fully agree with the importance of precise details such as how network, I/O, storage, memory, and even other (software) resources are managed, these specific details are already captured high-level in the architecture and more details fall outside of the scope of this work. 
The components \texttt{R3}, \texttt{R4} and \texttt{R5} capture currently the management and scheduling of these resources. 

Similarly to our treatment of I/O, representing the Broker component as a stub in the reference architecture could pose a threat.
As indicated in Section~\ref{sec:model:scheduling}, we support the broker stub, but a full brokering system can be complex, and that would significantly imbalance the reference architecture.
As such, we treat the broker as any of the important scheduling stages, and not as a more complex system.

Another possible threat is {\it the lack of a closed-form mathematical model} supporting this work.
While useful, we argue such a model will be very hard to construct in the face of the curse of dimensionality.
Thus, we consider constructing such a model future work for the community, albeit, we warn it may not be possible to subsume the entire complexity of datacenter scheduling in a tractable, closed-form formula.

\subsection{Feasibility of Mapping New Schedulers}\label{sec:mapping:repro}\label{sec:threats:mapping}

\begin{figure}[t]
	\centering
    \includegraphics[width=\columnwidth]{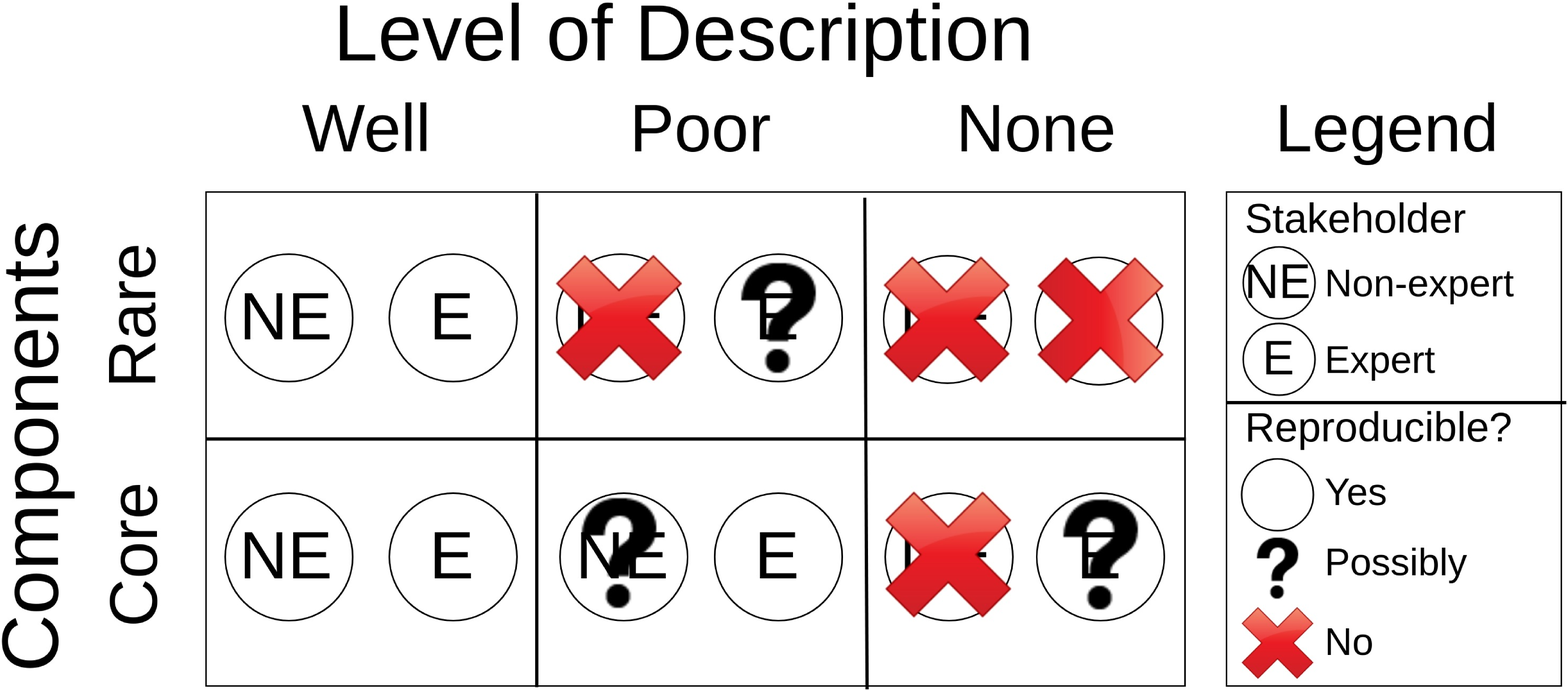}
    \caption{A matrix depicting the ability of non-expert and expert users to reconstruct components.}
    \label{fig:knowledge-matrix}
\end{figure}

Through a thought experiment, we analyze in this section the feasibility of mapping new datacenter schedulers to our reference architecture. That this is feasible is important, because it illustrates the validity (accuracy) and ease of use of our reference architecture. Conducting real-world experiments requires access to a sufficient fraction of the community's experts and novices, whose participation in a large-scale study depends on the popularity of the reference architecture---a circular problem we have begun addressing through direct contact, but which falls outside the scope of this work.

To understand the feasibility of mapping, we first define categories for how common the components are, how well the components are described, and for how well trained the analysts are. These categories are:
(1) As quantified in Section~\ref{sec:validation:overview}, some component are highly present in descriptions of schedulers; without imposing an artificial threshold, we assume some scheduling components are very commonly described in practice and known by all practitioners, and thus constitute {\it core} components. Conversely, the other components are relatively {\it rare}, and are not commonly known by practitioners.
(2) As described in Section~\ref{sec:validation:process}, components can offer full, partial, and no matches; the matches are derived from these components being well-, poorly, and not described, respectively. 
(3) We focus on two extreme types of analysts using our reference architecture, either experts or non-experts.
We assume {\it experts} have in-depth knowledge of several schedulers common in datacenters, including both conceptual and technical forms.
{\it Non-experts} are unfamiliar with datacenter scheduling systems, yet are able to reason about and implement systems when provided written source designs and related material e.g., peer-reviewed articles and/or source-code.

We analyze through a though experiment the full matrix of combinations between component rarity and level of description, for each level of expertise. Figure~\ref{fig:knowledge-matrix} depicts an overview of the ability of (non-)expert analysts to map (reproduce) a component. We find the following important cases:
(1) Well-described components can be reconstructed by both expert and non-expert users, regardless of their rarity; even for the rarest of components, non-experts can simply follow the description.
(2) Core components that are also poorly described can often be reproduced by expert users, which are familiar with the general aspects missing from the description. Non-expert users could reproduce such components through guesswork, largely by looking at similar core components that are well-described.
(3) Rare components that are also poorly described could be reconstructed by experts through guessing or sheer experience. However, they cannot be reconstructed by non-experts, who lack knowledge and easily available alternatives cannot reconstruct the missing parts.
(4) Finally, components that are not described at all may be reproduced by experts if common, because experts will realize such components \textit{must} be present in a working design and, through sheer experience, devise an exact or alternative design. However, rare components are too specialized or uncommon to be accurately recreated even by experts, unless code or documentation become available, or a conversation with the developers can take place. Moreover, components with no description cannot be reproduced by non-expert users, who normally will not even realize the necessity of such components being present in the design. 

\section{Related Work}
\label{sec:related-work}\label{sec:related}

In this section, we survey three main classes of related work, in contrast to which we claim significant novelty in Section~\ref{sec:related}. We explain in the following our claims:

\subsubsection{vs. Scheduling architectures for datacenter-based environments} 
This work finds its primary inspiration in Schopf's multi-stage model of the grid scheduling process~\cite{Schopf2004}, and subsequent work in the Global Grid Forum (e.g.,~\cite{DBLP:conf/coregrid/GrimmeLPWYOWZ08}). 

We propose a conceptual model that significantly expands Schopf's model, with higher granularity of stages, support for advanced concepts typical in modern datacenters (e.g., the concept of workflows, operations such as replication, checkpointing, and migration, etc.) This work also conducts an empirical validation, and trace-based
experiments.

In survey work, Rodriguez et al.~\cite{Rodriguez2017} and Singh et al.~\cite{Singh2016} discuss the general scheduling process in datacenters and propose taxonomies of resource-management systems. 
Rodriguez et al. focus on broader architectures, in which schedulers play a coarse-grained, black-box role. 
Singh et al. cover the resource provisioning and scheduling process, but lack the fine granularity and advanced concepts of this work. 

\subsubsection{vs. Architectures of grid and cloud computing, and of big data systems} 
The architectures proposed at this level
are coarse-grained.
This holds for various architectures proposed for grids, e.g., in the seminal work of Foster et al.~\cite{DBLP:journals/ijhpca/FosterKT01} or by the GGF, and the reference architectures proposed by the National Institute of Standards and Technology (NIST) for cloud computing~\cite{NISTCloudRA} and for big data systems~\cite{NISTBigDataRA}. 
Our work complements these approaches, and 
fits within the general framework of the \texttt{ISO/IEC/IEEE 42010:2011} standard~\cite{ISOSoftArch}, 
as an ``architecture framework'' establishing the ``common practice'' of the field.

\subsubsection{vs. Architectures of large-scale software systems} 
Our reference architecture aligns with other conceptual models and architectures for (large-scale) software. As such, it builds upon the existing theory of software architecture design. Rozanski and Woods discuss common principles in software architecture~\cite{Rozanski2005} (see Section~\ref{sec:reference-architecture} for a discussion of how we use their principles). This reference architecture extends the ``Pipes and Filters'' model~\cite{Rozanski2005}.

Bass et al.~\cite{Bass2003} propose a taxonomy of abstractions of software architectures, from which our work matches {\it reference architectures}---``reference model mapped onto software elements (that cooperatively implement the functionality defined in the reference model) and the data flows between them''.

\section{Conclusion and Ongoing Work}
\label{sec:conclusion}

With cloud operations gaining importance across society, efficient management of datacenter workloads and resources is increasingly critical. 
The main component responsible for this is the scheduler, which allocates user workloads to resources. 
Although many scheduling systems already exist, their complex nature and diverse approaches makes it difficult to comprehend and compare them. 
To address this problem, we propose a reference architecture for datacenter scheduling.

Our reference architecture focuses on a conceptual workflow-based model, whose tasks and flows embody the processes for scheduling in datacenters. 
We validate this model by mapping to it fourteen existing schedulers, traditional and recent, sourced both from academia and industry.
We further give evidence of how the reference architecture can be used in practice, through static and dynamic analysis. 
Trace-based simulations indicate that the underspecification of components, which is common in published material about schedulers, is problematic for the reproducibility and interpretability of results. 
From an empirical, trace-based study of stage-complexity at runtime, we find that the separation into stages is meaningful for understanding the dynamic operation of schedulers, and underspecification of key stages can impact significantly online usability.
Overall, we conclude that our reference architecture offers a conceptual model that unifies scheduling approaches, is representative for current state-of-the-art, and can help understand datacenter schedulers. 

For the future, we plan to explore both the scientific and the engineering potential of this architecture. 
On the scientific side, we will extend the validation effort in this work to more schedulers and explore how emerging paradigms (e.g., IoT, edge, and serverless) can be reconciled with the architecture. 
We also aim to further explore the design space of scheduling with this model, through trace-based simulation and real-world experiments, including with a real-world full-featured scheduler (e.g., based on Kubernetes and Condor). 
Furthermore, we foresee our reference architecture to contribute to the design of taxonomies for scheduling systems, enabling further understanding and design studies.
From an engineering perspective, we see significant potential for the development, tuning, and analysis of schedulers with this model. 
For reproducibility considerations, we will release both open-source code and open-access data, see details in Appendix~\ref{sec:artifact}.

Long-term, we envision and are currently developing the tools supporting a global competition for scheduling in datacenters, where participants can develop schedulers to address yearly scenarios~\cite{opendc}.
Scheduler submissions run against the workloads and resources specified by each scenario, in reproducible experiments.
Finally, the results are analyzed and made public, with participants getting extensive reports of their experimental evaluation and public praise. 
We see this approach as facilitating direct practical experience with scheduler development, for participants of varying backgrounds.

Considering techniques derived from social sciences, which are currently beginning to be used to expand the capabilities of the software engineering community, we also envision conducting and analyzing interviews with the original authors of the schedulers used in this work.

\section*{Acknowledgments}
This work was supported by Dutch projects Vidi MagnaData and Commit.
We thank the members of the AtLarge team who helped in the process of creating this publication: Alexandru Uta, Erwin van Eyk, and Lucian Toader.

\balance

\bibliographystyle{IEEEtran}
\bibliography{bibliography_cut_beautified}

\pagebreak

\nobalance 

\appendices
\section{Artifact Description: ``A Reference Architecture for Datacenter Scheduling: Design, Validation, and Experiments''}
\label{sec:artifact}

%%%%%%%%%%%%%%%%%%%%%%%%%%%%%%%%%%%%%%%%%%%%%%%%%%%%%%%%%%%%%%%%%%%%%
\subsection{Abstract}

This description contains the information needed to reproduce the experiments outlined in Section~\ref{sec:experiments}. We have reproduced our work on MacOS El Capitan, Windows 10, and the lightweight Alpine Linux 3.8.

We have released our software as open-source, as part of the SPEC Research community-driven, open-source OpenDC platform~\cite{opendc}.
We have also released all the experimental data produced in this work, through the Zenodo Open Science platform\footnote{\url{https://zenodo.org/record/1343629}}.

%%%%%%%%%%%%%%%%%%%%%%%%%%%%%%%%%%%%%%%%%%%%%%%%%%%%%%%%%%%%%%%%%%%%%
\subsection{Description}

\subsubsection{Check-list (artifact meta information)}

{\small
\begin{itemize}
  \item {\bf Algorithm: } Resource management and scheduling system, with policy configurations for stages
  \item {\bf Program: } Kotlin (JVM-based) codebase and dependencies
  \item {\bf Compilation: } Java Development Kit, Kotlin, Gradle
  \item {\bf Data set: } free open-access data collected by the community into the Grid Workload Archive traces~\cite{DBLP:journals/fgcs/IosupLJADWE08}
  \item {\bf Run-time environment: } Java Virtual Machine
  \item {\bf Output: } Stage execution times and task turnaround times
  \item {\bf Experiment workflow: } Docker
  \item {\bf Experiment customization: } Choice of workload, scheduler configurations, and number of repetitions
  \item {\bf Publicly available?: } Yes, on GitHub and Zenodo
\end{itemize}
}

\subsubsection{How software can be obtained (if available)}
The software can be obtained from the {\tt sc18} release of the OpenDC simulator GitHub repository\footnote{\url{https://github.com/atlarge-research/opendc-simulator/releases/tag/sc18}}. We follow the approach of OpenDC and offer next to source-code a container-based deployment approach, based on Docker.

\subsubsection{Hardware dependencies}
The software runs on CPU architectures officially supported by Docker and the JVM. We recommend the \textit{x86-64} architecture as other architectures may require building images locally. 

\subsubsection{Software dependencies}
The software requires Docker to run. The other dependencies are encapsulated within the Docker deployment.

\subsubsection{Datasets}
Traces from the Grid Workload Archive~\cite{DBLP:journals/fgcs/IosupLJADWE08} are used for the experiments. A copy of the traces is also included in the repository.

%%%%%%%%%%%%%%%%%%%%%%%%%%%%%%%%%%%%%%%%%%%%%%%%%%%%%%%%%%%%%%%%%%%%%
\subsection{Installation}
The experiments require a running installation of the Docker environment\footnote{\url{https://www.docker.com/}}.

%%%%%%%%%%%%%%%%%%%%%%%%%%%%%%%%%%%%%%%%%%%%%%%%%%%%%%%%%%%%%%%%%%%%%
\subsection{Experiment workflow}
The experiments take approximately 8 hours to complete.
The following files (available in the data archive on Zenodo) need to be present in the current working directory for either of the two experiments to start:
\begin{enumerate}
\item \textbf{/setup.json} - JSON file describing the topology of the datacenter
\item \textbf{/askalon\_workload\_ee.gwf} - GWF file containing the trace for the \wlaskalon{} workload
\item \textbf{/chronos\_exp\_noscaler\_ca.gwf} - GWF file containing the trace for the \wlchronos{} workload
\end{enumerate}

The \wlaskalon{} experiments can be started as follows:
\begin{verbatim}
$ docker run -it --rm \
   -v $(pwd):/home/gradle/simulator/data \
   atlargeresearch/sc18-experiment-runner \
   -r 32 -w 4 \
   -s data/setup.json \
   data/askalon_workload_ee.gwf
\end{verbatim}

After the \wlaskalon{} experiments have been finished, the \wlchronos{}
experiments can be started in a similar fashion.
The result files of any previous experiment runs should be backed up in a different location before running this command, as it will overwrite the result files.
\begin{verbatim}
$ docker run -it --rm \
   -v $(pwd):/home/gradle/simulator/data \
   atlargeresearch/sc18-experiment-runner \
   -r 32 -w 4 \
   -s data/setup.json \
   data/chronos_exp_noscaler_ce.gwf
\end{verbatim}

For a more in-depth guide we refer to the \texttt{README.md} file associated with the software artifacts.

%%%%%%%%%%%%%%%%%%%%%%%%%%%%%%%%%%%%%%%%%%%%%%%%%%%%%%%%%%%%%%%%%%%%%
\subsection{Evaluation and expected result}
The output can be found as comma-separated values files in the \texttt{data/} directory of the repository after the experiments have finished.
The data files contain the duration of each stage per scheduling cycle and the metrics of the tasks run during simulation (waiting, execution, turnaround time).

%%%%%%%%%%%%%%%%%%%%%%%%%%%%%%%%%%%%%%%%%%%%%%%%%%%%%%%%%%%%%%%%%%%%%
\subsection{Experiment customization}
The experiment runner allows for customization of the experiments through the optional command line arguments listed below:
\begin{itemize}
 	\item \textbf{\texttt{-r}}, \textbf{\texttt{--repeat}} \\
    The number of times to repeat an experiment for each scheduler.
    \item \textbf{\texttt{-w}}, \textbf{\texttt{--warm-up}} \\
     The number of times to run a warm-up experiment before starting the recorded experiments for each scheduler.
    \item \textbf{\texttt{-p}}, \textbf{\texttt{--parallelism}} \\
     The number of experiments to run in parallel.
    \item \textbf{\texttt{--schedulers}} SCHEDULERS \\
 	The list of schedulers to test, separated by spaces.
\end{itemize}

%%%%%%%%%%%%%%%%%%%%%%%%%%%%%%%%%%%%%%%%%%%%%%%%%%%%%%%%%%%%%%%%%%%%%
\subsection{Notes}
The amounts of resources (e.g., clusters) in the datacenter can be set in \texttt{setup.json}. Each item represents the identifiers of the resource (here, CPU type) to use in the machine. The available CPU types are (1) Intel i7 (4 cores, 4100 MHz) and (2) Intel i5 (2 cores, 3500 MHz).
\newpage

\end{document}